\PassOptionsToPackage{table}{xcolor}
\documentclass[acmsmall]{acmart}
\usepackage[]{hyperref}
\usepackage{xurl}
\usepackage{enumitem}
\usepackage{acronym}
\usepackage{multirow}
\usepackage{xspace}
\usepackage{cleveref}
\usepackage[htt]{hyphenat}
\usepackage{booktabs}
\usepackage{graphics}
\usepackage{multicol}
\usepackage{tabularx}
\usepackage{diagbox}
\newcolumntype{L}{>{\raggedright\arraybackslash}X} 
\usepackage{array}
\usepackage[commandnameprefix=always]{changes}
\newcommand{\qppnet}{\texttt{QPP-Net}\xspace}
\newcommand{\postgres}{\texttt{Scaled Postgres}\xspace}
\newcommand{\postgresx}{\texttt{Scaled PG10}\xspace}
\newcommand{\postgresxvi}{\texttt{Scaled PG16}\xspace}
\newcommand{\mscn}{\texttt{MSCN}\xspace}
\newcommand{\etoe}{\texttt{End-To-End}\xspace}
\newcommand{\dace}{\texttt{DACE}\xspace}
\newcommand{\zeroshot}{\texttt{Zero-Shot}\xspace}
\newcommand{\flatvector}{\texttt{Flat Vector}\xspace}
\newcommand{\queryformer}{\texttt{QueryFormer}\xspace}
\newcommand{\lcm}{\ac{lcm}\xspace}
\newcommand{\lcms}{\acp{lcm}\xspace}
\newcommand*\circles[1]{\raisebox{.5pt}{\textcircled{\raisebox{-.8pt}{#1}}}}

\theoremstyle{definition}
\newtheorem{definition}{Definition}
\makeatletter
\def\thm@space@setup{%
  \thm@preskip=0pt   
  \thm@postskip=0pt  
}
\makeatother

\acrodef{lcm}[LCM]{\emph{Learned Cost Model}}
\acrodef{mlp}[MLP]{\emph{Multi-Layer Perceptron}}
\setcopyright{acmlicensed}
\copyrightyear{2025}
\acmYear{2025}
\acmDOI{XXXXXXX.XXXXXXX}
\acmJournal{PACMMOD}
\acmVolume{X}
\acmNumber{X}
\acmArticle{XX}
\acmMonth{02}

\author{Roman Heinrich}
\orcid{0000-0001-7321-9562}
\affiliation{%
 \institution{Technical University of Darmstadt \& DFKI Darmstadt}
 \city{Darmstadt}
 \country{Germany}}

\author{Manisha Luthra}
\orcid{0000-0002-3788-6664}
\affiliation{%
 \institution{Technical University of Darmstadt \& DFKI Darmstadt}
 \city{Darmstadt}
 \country{Germany}}
 
\author{Johannes Wehrstein}
\orcid{0000-0002-7152-8959}
\affiliation{%
 \institution{Technical University of Darmstadt}
 \city{Darmstadt}
 \country{Germany}}
 
\author{Harald Kornmayer}
\orcid{0000-0002-8878-7623}
\affiliation{%
 \institution{Duale Hochschule Baden-Württemberg (DHBW) Mannheim}
 \city{Mannheim}
 \country{Germany}}
 
\author{Carsten Binnig}
\orcid{0000-0002-2744-7836}
\affiliation{%
 \institution{Technical University of Darmstadt \& DFKI Darmstadt}
 \city{Darmstadt}
 \country{Germany}}


\begin{document}
\title{How Good are Learned Cost Models, Really?\\Insights from Query Optimization Tasks}
\begin{abstract}
Traditionally, query optimizers rely on cost models to choose the best execution plan from several candidates, making precise cost estimates critical for efficient query execution. 
In recent years, cost models based on machine learning have been proposed to overcome the weaknesses of traditional cost models. 
While these models have been shown to provide better prediction accuracy, only limited efforts have been made to investigate how well \lcms actually perform in query optimization and how they affect overall query performance. 
In this paper, we address this by a systematic study evaluating \lcms on three of the core query optimization tasks: \emph{join ordering}, \emph{access path selection}, and \emph{physical operator selection}. 
In our study, we compare seven state-of-the-art \lcms to a traditional cost model and, surprisingly, find that the traditional model often still outperforms \lcms in these tasks.
We conclude by highlighting major takeaways and recommendations to guide future research toward making \lcms more effective for query optimization.
\end{abstract}
\maketitle

\section{Introduction} \label{sec:introduction}
\noindent\textbf{Cost Estimation is Crucial for Databases.}
Accurate cost prediction of a query plan is essential for optimizing the query performance in a database.
During query optimization, estimated costs of various candidate plans guide the plan selection for execution \cite{leis_how_2015}.
Consequently, accurate cost estimates are pivotal in query optimization.
In the worst case, incorrect cost estimates can lead to the selection of highly unfavorable plans with a runtime that is several factors higher than the optimal one.
However, it is well known that providing accurate cost estimates is difficult and inaccurate estimates influence the results of finding an optimal plan significantly \cite{lan2021survey, heinrich2024}.

\noindent\textbf{The Need for Accurate Cost Models.}
In recent decades, classical rule- or heuristic-based cost estimation methods have been the backbone for query optimization in commercial DBMSs.
However, since they rely on heuristics and simple analytical models, traditional methods struggle with accuracy and provide estimates that often deviate by orders of magnitude from the actual execution costs \cite{leis_how_2015}, leading to sub-optimal query execution plans with prolonged runtimes.
Thus, much research has been dedicated to making cost models more precise and improving overall performance \cite{karampaglis2014, he2005}.

\noindent\textbf{The Rise of Learned Cost Models.} 
Supported by the high potential of Machine Learning (ML), many \acfp{lcm} have been proposed over the last years \cite{hilprecht2022, ganapathi2009, akdere2012, kipf2019, marcus2019, zibo_liang_dace_2024, yang2023, zhao2022, li_learned_2024, chang2024, lu2022, duggan2011, heinrich2024, zhou2020, wu2022, agnihotri2024}.
These models typically leverage actual costs from executing training queries to learn patterns and predict the execution cost of new queries.
The main promise of \lcms lies in the fact that they can better capture complex data distributions and inter-dependencies with workloads and thus potentially lead to more efficient query processing and shorter query runtimes.
As a result, many recent papers about \lcms \cite{zibo_liang_dace_2024, sun2019, hilprecht2022} show that they can significantly outperform classical cost models that are employed in traditional database systems such as PostgreSQL in terms of cost prediction.

\begin{figure}
    \centering
    \includegraphics[width=0.8\linewidth]{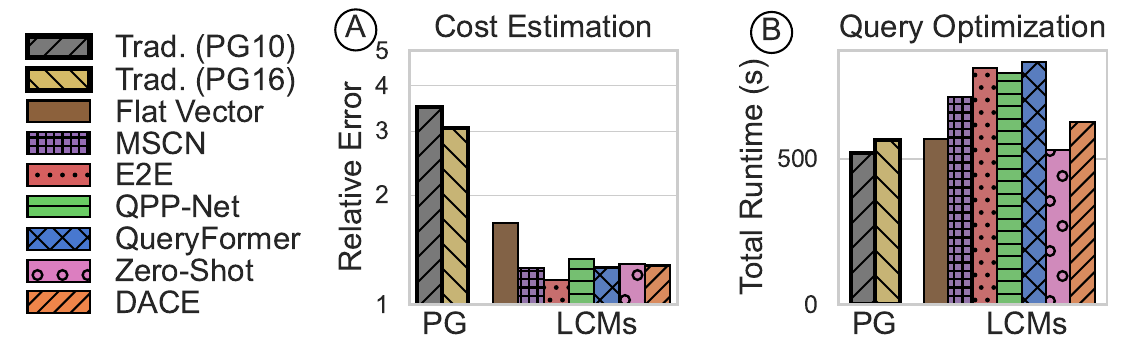}
    \caption{\circles{A} \lcms outperform traditional approaches in terms of cost estimation on an unseen IMDB dataset. 
    \circles{B} However, when optimizing a workload (JOB-Light) for join order, the traditional PostgreSQL (PG) model still performs best.}
    \label{fig:motivating_plot}
\end{figure}

\noindent\textbf{Do \lcms Really Help in Query Optimization?}
However, while \lcms have led to improvements in accuracy, a core question is if they lead to improvements for query optimization and to what extent. 
Surprisingly, most existing evaluations primarily focus on the accuracy of the cost estimation task and largely neglect a deeper analysis of how these cost models actually improve query optimization \cite{hilprecht2022, ganapathi2009, akdere2012, kipf2019, marcus2019, zibo_liang_dace_2024, sun2019, yang2023, zhao2022, lu2022, zhou2020}.
In this paper‚ we argue that accuracy alone is not meaningful, as it cannot reflect important tasks in query optimization, such as the ranking and selection of plans.
We thus aim to close this gap and conduct a systematic study to assess \textit{how good learned cost models really are for query optimization?}
Unfortunately, the results of our study are rather grim, as shown by \Cref{fig:motivating_plot}, which highlights some of the findings of our study.
In \Cref{fig:motivating_plot} \circles{A}, we compare the accuracy of a broad spectrum of recent \lcms. Here, in terms of prediction error, the traditional approach PostgreSQL (PG, black bar) is outperformed by all \lcm competitors (colored bars) on the IMDB dataset\footnote{We report the median Q-error as standard metric for cost models. It defines the relative deviation of the predicted from the actual cost. A perfect prediction has a Q-error of $1$. See \Cref{subsec:setup} for more details on the setup.}
However, the picture is very different when using the cost models for finding optimal join orders on the JOB-Light benchmark \cite{kipf2019}. 
\Cref{fig:motivating_plot} \circles{B} shows that the total query runtime is \textit{not} improving when using \lcms for join ordering.
Here, \textit{none} of the \lcms is able to provide better selections than PostgreSQL, resulting in a higher total runtime of selected plans on the JOB-Light benchmark of up to 832s, whereas PostgreSQL achieves 510s. 

\noindent\textbf{A Novel Evaluation Study.}
From these results, it becomes clear that focusing alone on the accuracy of cost estimation is not sufficient.
As such, in this paper, we provide a systematic study to shed some light on the question of why \lcms fail to enable better optimizer decisions.
To answer this question, we have chosen \textit{three} of the most important tasks of query optimization (\textit{join ordering}, \textit{access path selection}, and \textit{physical operator selection}) and analyze whether or not \lcms are able to improve plan selection.
As a main contribution, we evaluate a set of recent \lcms that cover a broad spectrum of approaches proposed in the literature and compare their impact on query optimization against the traditional cost model of PostgreSQL.
We suggest a task-specific, fine-grained evaluation strategy for each downstream task that goes beyond prediction accuracy, assessing how \lcms affect query optimization.

\noindent\textbf{Key Insights of Our Study.}
Our evaluation reveals three key insights that we summarize in the following:
\begin{enumerate}[leftmargin=*, nosep]
    \item \textit{High accuracy in cost is not sufficient}: Across all analyzed query optimization tasks, we observed that it is insufficient to focus solely on the prediction accuracy of plan costs.
    Instead, \lcms need to fulfill other properties such as reliable ranking and selection of plans.
    Moreover, existing \lcms majorly only optimize for median prediction errors while possessing high errors in the tail, leading to large over- and underestimation and making \lcms prone to select non-optimal plans.
    \item \textit{Training data matters}:
    As \lcms are typically trained on queries that have been pre-optimized by a traditional query optimizer, their training data is biased towards near-optimal plans.
    However, during query optimization, \lcms must predict costs for both optimal and non-optimal plans. 
    Moreover, training data quality can also introduce other biases, especially when timeouts are used during query execution for training data collection, distorting the \lcms understanding of ``bad'' plans. 
    For instance, query plans with nested-loop joins may often timeout before completion and thus, only the cases where nested-loop joins are beneficial are included in the training data.
    \item \textit{Don't throw away, what we know}:
    Traditional cost models often deviate significantly from actual costs in their estimates. 
    However, they incorporate extensive expert knowledge based on years of experience.
In our paper, we found that using their estimates as input to \lcms is highly beneficial as it significantly improves the cost estimates for query optimization tasks.
\end{enumerate}
\noindent\textbf{Consequences for \lcms.} 
Overall, we believe that the results of our study can guide future research and development efforts toward more reliable ML-based cost estimation that makes informed decisions about query optimization tasks. 
We discuss some directions based on the evidence of this paper that will help to provide \lcms, which actually benefits query optimization.
Furthermore, to enable the research community to build on our results, we made the source code, models, and all the evaluation data publicly available\footnote{
\label{code_link}\textbf{Source Code}: \url{https://github.com/DataManagementLab/lcm-eval};\\
\label{data_link}\textbf{Experimental data \& trained models}:
\url{https://osf.io/rb5tn/}}

\noindent\textbf{Outline.}
We first provide a background in cost estimation in \Cref{sec:background}.
Next, we present our evaluation methodology in \Cref{sec:methodology}, including a taxonomy of recent \lcms.
In the subsequent sections, we then evaluate the downstream tasks of join ordering (\Cref{sec:join_order}), access path selection (\Cref{sec:access_path_selection}), and physical operator selection (\Cref{sec:physical_operator}).
We provide our recommendations for \lcms in \Cref{sec:lessons} and summarize this paper in \Cref{sec:conclusion}.
\section{Background of Cost Estimation} \label{sec:background}
This section first gives a brief overview of classical and learned cost estimation. 
Afterwards, we describe the learning procedure of \lcms and provide a taxonomy that guides our selection of recent \lcms for this study in \Cref{sec:methodology}.

\subsection{Traditional \& Learned Cost Estimation}
\textbf{Traditional Cost Estimation.} 
Precise cost estimates for different plan candidates in a database are crucial for the query optimizer to select optimal plans from a large search space.
Thus, a lot of engineering effort has been spent since the beginning of database development to estimate the execution costs of a query plan.
Most database systems such as MySQL \cite{widenius2002}, Oracle, PostgreSQL, or System R \cite{astrahan1976} use hand-crafted cost models to reason about the execution costs of a query plan.
These models typically provide a cost function for each physical operator in a query plan that estimates its runtime costs according to CPU usage, I/O operations, memory consumption, expected tuples, and random or sequential page accesses.
However, due to the wide variety of data, queries, and data layouts, traditional cost models need to make simplifying assumptions (e.g., independence of attributes).
These often lead to incorrect predictions of the execution cost. 
Consequently, the query optimizer makes sub-optimal decisions that degrade the query performance by increasing its runtime \cite{leis_how_2015}.

\noindent\textbf{Learned Cost Estimation.}
The need to improve prediction accuracy and the rise of machine learning motivated the idea of \lcms. 
The main idea is to approximate the complex cost functions with a learned model.
Generally, a typical model learns from previous query executions to predict execution costs like runtime.
In contrast to traditional cost models, the promise of \lcms is that they can better learn arbitrarily complex functions.
Thus, improved prediction accuracy can be expected in contrast to traditional approaches based on simplifying assumptions.
Overall, the higher accuracy is expected to lead to a selection of query plans with improved query performance.

\subsection{Learning Procedure of \lcms}
\begin{figure*}
    \centering
    \includegraphics[width=\linewidth]{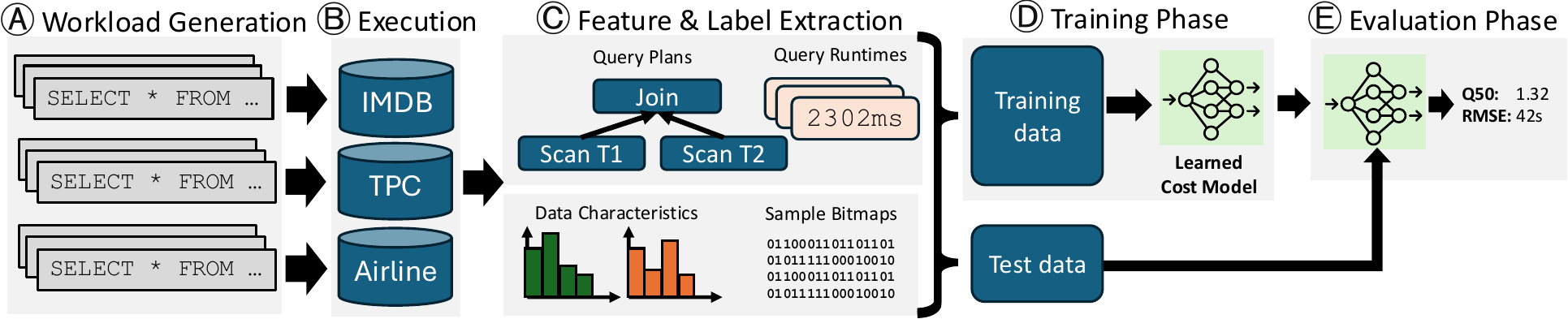}
    \caption{
    Learning procedure of \lcms. 
    \circles{\textsc{A}} Generation of synthetic training queries. \circles{\textsc{B}} Query execution on training databases. 
    \circles{\textsc{C}} Feature (query plans, data characteristics, and sample bitmaps) and label (query runtimes) extraction to generate the training and test dataset. 
    \circles{\textsc{D}} Training of the \lcm with supervised learning. 
    \circles{\textsc{E}} Evaluation of the \lcm against unseen test data.}
    \label{fig:learning_procedure}
\end{figure*}

For our study, we look at effects that also result from the learning procedure of \lcms.
As such, we briefly review the traditional procedure as depicted in \Cref{fig:learning_procedure} to provide the necessary background:
\circles{A}~At first, a workload generator is used to create a large set of randomized, synthetic SQL-Strings that involve a variety of representative query properties such as filter predicates, joins, or aggregation types.
\circles{B}~These queries are executedses (e.g., an airline or movie database) to collect the actual costs of queries.
An important aspect here is that training procedures of many \lcms leads to biases in the dataset due to timeouts and pre-optimized queries, as discussed later.
\circles{C}~Next, various information is extracted from the workload execution.
Most importantly, the physical query plans are extracted, which serve as input to cost models.
In addition to physical plans, \lcms require different information, such as data characteristics like histograms or sample bitmaps.
\circles{D}+\circles{E}~Finally, the workload (i.e., plans and runtime) is then split for training and testing the \lcms. 

\subsection{Taxonomy of \lcms} \label{subsec:taxonomy}
\lcms developed in the last years differ in various dimensions.
This section provides a brief taxonomy of recent \lcms to structure the different methodological approaches. 
This taxonomy will guide the selection of \lcms that we use in this study and ensure that we cover the different methodologies to analyze how they affect the ability of \lcms to support query optimization.

\noindent\textbf{Input Features.}
The first crucial dimension is the input features that a \lcm learns from.
The input features are extracted from the executed workloads (cf. \Cref{fig:learning_procedure}\circles{C}).
The query plan and the underlying data distribution need to be modeled so that a \lcm is informed to make reasonable predictions about the execution costs, which in turn affects query optimization, as we will show.
However, \lcms make use of different information for cost estimation.

\begin{enumerate}[leftmargin=*, nosep]
\item \textbf{SQL-String vs. Query Plans}: 
Some of the first models rely on the SQL string to describe a query, as it gives insights about the tables, predicates, and joins. 
However, details of the execution plan, such as physical operators or the order of joins, are not described there. 
Thus, most \lcms utilize the physical query plan, which includes the operators (e.g., scans, joins) and physical operator types  (e.g., nested loop vs. hash join).
As we will see later, this is fundamental for query optimization.

\item \textbf{Cardinalities}:
Intermediate cardinalities are an important input signal for the overall cost of a plan as they denote the number of tuples an operator needs to process \cite{leis_how_2015}.
Thus, many \lcms leverage intermediate cardinalities as input features, which are either annotated by the databases' cardinality estimator or obtained through an additional learned estimator from related work \cite{hilprecht2020deepdb, kipf2019, yang2020}.
While some \lcms also ignore cardinalities as input for cost prediction, we show in our study that they, in fact, improve the usefulness of cost estimates from \lcms for various query optimization tasks.

\item\textbf{Data Distribution}:
Another helpful factor in estimating cost is understanding the data distribution in the base tables, especially if no cardinalities are used.
For instance, the fact of how many distinct values exist in a column might influence the efficiency of physical operators (e.g., hash join). 
As such, some \lcms use data distribution represented as database statistics and histograms or sample bitmaps (which we explain later) from the base tables as inputs.
However, as we will show in our study, their effect on query optimization tasks remains unclear.

\item \textbf{Cost Estimates}: 
Finally, some of the most recent \lcms even leverage the cost estimates provided by a classical cost estimator as an input feature, which serves as a strong input signal.
This idea renders these \lcms to \textit{hybrid} as they combine a traditional cost model with a learned approach.
The study shows that this provides significant benefits.
\end{enumerate}

\noindent \textbf{Query Representation.}
Many \lcms use model architectures use a graph-based representation to encode query plans as input to the models\footnote{The graph-based representation of the queries refers to the fact whether a model leverages the query graph structure and not to the model learning architecture itself.}.
These approaches thus explicitly leverage information about the order (parent-child relationships) of operators in plans.
However, other \lcms \cite{kipf2019, akdere2012} represent a query plan (or the SQL string) as a flat vector of fixed size without modeling the operator dependencies, which we refer to as \textit{flat} representation in this paper.
While intuitively, capturing the structure and not using a flat representation should be beneficial for \lcms, the results of using graph structure in this study are not that clear. 

\noindent \textbf{Database Dependency.}
Furthermore, an important aspect is whether \lcms can generalize to unseen databases (i.e., a new set of tables) or not.
\textit{Database-agnostic} \lcms were designed \cite{hilprecht2022, zibo_liang_dace_2024} to enable cost predictions for unseen databases that were not part of the training data.
This approach has the advantage of directly providing results without requiring database-specific training data. 
In contrast, \textit{database-specific} \cite{sun2019, zhao2022, marcus2019} models cannot generalize for unknown databases.
For this study, an interesting question is if one of these classes is better suited to support query optimization tasks as database-specific can better adapt to one single database while database-agnostic models can generalize better.

\noindent \textbf{Model Architecture.}
Finally, the presented \lcms differ largely in their learning approach.
Various learning architectures were proposed, including decision trees, tree-structured neural networks, neural units, graph neural networks, and transformer architectures.
While different architectures show different results on the cost estimation tasks, it is still open to see which architecture provides the best results for query optimization.
\section{Evaluation Methodology} \label{sec:methodology}
In this section, we discuss the selection of \lcms for this study and then explain our evaluation strategy.
Afterwards, we discuss which downstream tasks we include in our study and why we select those tasks.
Finally, we explain the experimental setup. 
\begin{table*}
\resizebox{\linewidth}{!}{
\begin{tabular}{lr|cccccc|c|c|c}
\multicolumn{2}{c|}{} & \multicolumn{6}{c|}{\textbf{Input Features}} &  &  &  \\
\multicolumn{2}{c|}{\textbf{Model}} & \begin{tabular}[c]{@{}c@{}}SQL\\ String\end{tabular} & \begin{tabular}[c]{@{}c@{}}Physical\\ Plan\end{tabular} & Cardinalities & \begin{tabular}[c]{@{}c@{}}DB Cost \\ Estimates\end{tabular} & \begin{tabular}[c]{@{}c@{}}DB\\ Statistics\end{tabular} & \begin{tabular}[c]{@{}c@{}}Sample \\ Bitmaps\end{tabular} & \textbf{\begin{tabular}[c]{@{}c@{}}Query \\ Representation\end{tabular}} & \textbf{\begin{tabular}[c]{@{}c@{}}Database-\\ Dependency\end{tabular}} & \textbf{\begin{tabular}[c]{@{}c@{}}Model\\ Architecture\end{tabular}} \\ \hline
\cellcolor[HTML]{8c613c}\textcolor{white}{\textbf{\flatvector}} & \cellcolor[HTML]{8c613c}\textcolor{white}{\textbf{\cite{ganapathi2009}}} &  & \checkmark & \checkmark &  &  &  & Flat & DB-agnostic & Regression Tree \\ \hline
\cellcolor[HTML]{956cb4}\textcolor{white}{\textbf{\mscn}} & \cellcolor[HTML]{956cb4}\textcolor{white}{\textbf{\cite{kipf2019}}} & \checkmark &  &  &  &  & \checkmark & Flat & DB-specific & Deep Sets \\ \hline
\cellcolor[HTML]{d65f5f}\textcolor{white}{\textbf{\etoe}} & \cellcolor[HTML]{d65f5f}\textcolor{white}{\textbf{\cite{sun2019}}} &  & \checkmark &  &  & \checkmark & \checkmark & Graph & DB-specific & Tree Structured NN \\ \hline
\cellcolor[HTML]{6acc64}\textcolor{white}{\textbf{\qppnet}} & \cellcolor[HTML]{6acc64}\textcolor{white}{\textbf{\cite{marcus2019}}} &  & \checkmark & \checkmark & \checkmark & & \checkmark & Graph & DB-specific & Neural Units \\ \hline
\cellcolor[HTML]{4878d0}\textcolor{white}{\textbf{\queryformer}} & \cellcolor[HTML]{4878d0}\textcolor{white}{\textbf{\cite{zhao2022}}} &  & \checkmark &  &  & \checkmark & \checkmark & Graph & DB-specific & Transformer \\ \hline
\cellcolor[HTML]{dc7ec0}\textcolor{white}{\textbf{\zeroshot}} & \cellcolor[HTML]{dc7ec0}\textcolor{white}{\textbf{\cite{hilprecht2022}}} &  & \checkmark & \checkmark &  & \checkmark &  & Graph & DB-agnostic & Graph Neural Networks \\ \hline
\cellcolor[HTML]{ee864a}\textcolor{white}{\textbf{\dace}} & \cellcolor[HTML]{ee864a}\textcolor{white}{\textbf{\cite{zibo_liang_dace_2024}}} &  & \checkmark & \checkmark & \checkmark &  &  & Graph & DB-agnostic & Transformer \\
\end{tabular}
}
\caption{Selection and main dimensions of \lcms for our study}
\label{tab:taxonomy}
\end{table*}

\subsection{Selection of \lcms}
Existing \lcms differ in various dimensions, which might affect query optimization.
For this paper, we thus select a representative, broad set of recent state-of-the-art \lcms covering various approaches as shown in \Cref{tab:taxonomy}.
Moreover, we focus on models where artifacts were available to make the results of this study reproducible. 
Below, we discuss the model selection briefly.

\begin{enumerate}[leftmargin=*, nosep]
    \item \flatvector \cite{ganapathi2009}:
    This is one of the earliest approaches published more than 15 years ago and serves as a simple baseline for more recent and complex approaches.
    This paper aims to represent the physical query plan as a fixed-size vector with one entry per operator type (e.g., hash join, nested loop join, sort-based aggregate, hash aggregate). 
    Each entry then contains the sum of the intermediate cardinalities per operator type.
    To predict the runtime of a plan, a state-of-the-art regression model LightGBM \cite{lightgbm} is trained, which uses such a flat vector as input.

    \item \mscn \cite{kipf2019}:
    As a second model, we choose \mscn as one of the first models of the more recent generation.
    While initially developed for cardinality prediction, the model has also been used for cost estimation \cite{hilprecht2022, yang2023, sun2019}. 
    We include this model as the only one that uses the SQL-String as input instead of the physical query plan.
    To represent the SQL query, \mscn uses one-hot encoded flat feature vectors to describe tables, join conditions, and predicates used in a query.
    Moreover, to learn from the data distribution, \mscn uses \textit{sample bitmaps} as model input, which indicates for the filter conditions of a given query which rows of the base tables qualify for selection.

    \item \etoe \cite{sun2019}:
    As the third approach, we select \etoe, as this was the first proposed \lcm which explicitly models the plan structure. 
    As a model architecture, a tree-structured neural network is used for which it combines different \ac{mlp} for encoding input features and aggregating them over the query graph. 

    \item \qppnet \cite{marcus2019}:
    This approach is also aware of the plan structure but uses a more modular approach of so-called \textit{neural units}, which are \ac{mlp}, trained per operator type (e.g., one for hash joins, one for nested loop joins).
    Each neural unit considers a set of operator-related features and predicts a per-operator runtime and hidden states, passed along the query graph as input signal for the following \ac{mlp}. 
    In addition, it learns from estimated cardinalities and estimated operator costs.

\item \queryformer \cite{zhao2022}:
    Different from \qppnet and \etoe, which are based on simple \ac{mlp}, this model employs a learning architecture using \textit{transformers} to estimate query costs. 
    Most importantly, it introduces a tree-based self-attention mechanism, which is designed to learn from long query plans with many operators.
    Furthermore, it also incorporates bitmap samples and a richer feature set compared to other \lcms, including histograms on base tables.
    
\item \zeroshot\cite{hilprecht2022}:
    This is the first \emph{database-agnostic} model proposed that can generalize across databases.
    To achieve this, it learns from so-called transferable features such as table size and does not encode database-specific features (e.g., attribute \& table names) as the models did before. 
    As a model architecture, variations of \textit{graph neural networks} are used. 

\item \dace \cite{zibo_liang_dace_2024}:
    Finally, \dace represents one of the most recent models, which combines transformers with a database-agnostic approach. 
    As it is based on transformers, it uses self-attention \cite{vaswani2017} and \textit{tree-structured attention} mechanisms.
    In contrast to other models, it uses a reduced feature set, mainly learning from the operator tree and the cost estimates provided by a traditional cost model (i.e., PostgreSQL cost). 
\end{enumerate}

\subsection{Tasks of Query Optimization}
In this study, we evaluate the selected \lcms against \textit{three} query optimization tasks that each rely on precise cost estimates but stress different abilities of cost models for plan selection.

\noindent\textbf{Join Ordering.}
The join order describes the sequence in which base tables are joined in a query plan.
It is decisive for the query runtime as joins are the most expensive operations in a query plan, and a sub-optimal join order can significantly increase the runtime as the intermediate cardinalities explode.
To solve this task during query optimization, \lcms are combined with plan enumeration techniques such as dynamic programming to select the plan with the estimated lowest cost for execution.

\noindent\textbf{Access Path Selection.}
In addition to the join order, the correct choice of access path (index vs. table scan) is another decisive factor for the runtime of the final query plan \cite{selinger1979}. 
Using indexes like B-trees for accessing data in queries can massively accelerate access when selected correctly by a cost model.
However, the cost of index-based access vs. scan-based access depends on several factors, such as selectivity or data distribution.
\lcms must fundamentally understand these to provide reliable cost estimates for access paths.

\noindent\textbf{Physical Operator Selection.}
Another critical task for query optimization is to select a physical implementation of query operators.
For instance, most database systems support various join implementations like hash join, sort-merge join, or nested-loop join.
Again, the optimal selection of physical operators depends on several factors, such as intermediate sizes or the sortedness of data, which are not all explicitly included in many \lcms.
 
\subsection{Experimental Setup} \label{subsec:setup}
In the following, we explain the experimental setup of this paper that is common for all these tasks.

\noindent\textbf{Training and Evaluation Data.}
To train and test the selected \lcms, we leverage the benchmark proposed by \cite{hilprecht2022} that contains 20 real-world databases and a workload generator that provides SPJA-queries supported by all presented \lcms.
The generator reflects the state-of-the-art workload generation used by recent \lcms to create training queries.
We generate and execute 10,000 queries per database to learn from a broad range of query patterns and set the timeout to 30s, which is sufficient to show the effects of selecting good plans.
As we will demonstrate, this runtime can show already significant trade-offs for query optimization.
Moreover, this is a realistic scenario for cloud providers \cite{renen2023, renen2024}.
We use the same dataset and workloads for all cost models. 
For the execution of queries, we use PostgreSQL v10.23 on bare-metal instances of type \texttt{c8220} on CloudLab as an academic resource \cite{duplyakin2019} to generate both our training and evaluation data.
Each query is executed three times, and the runtime is averaged for a stable ground truth.
For some experiments in our study, we enforce the plan selection with \texttt{pg\_hint\_plan}.
We make all models, evaluation results and training data publicly available\textsuperscript{\ref{data_link}}.

\noindent\textbf{Model Training.}
For the model training, we need to differentiate between database-specific and database-agnostic \lcms:

\textit{Database-specific} models are trained for a specific database (i.e., set of tables).
For these models, we ran 10,000 queries on a single database and divided them into a training, evaluation, and test set (80\%, 10\%, 10\%).
To ensure that database-specific models are sufficiently trained, we provided all of them with 10.000 training queries and validated that adding more training data does not improve the quality of their predictions.

\textit{Database-agnostic} models are trained across various databases. 
Thus, they typically require more training data but then generalize out-of-the-box to unseen datasets.
We trained all database-agnostic \lcms on 19 training databases, each with 5,000 queries, and evaluated on an unseen target database, according to the strategies proposed in \cite{hilprecht2022, zibo_liang_dace_2024}.
Training with more training data also did not yield significant improvements in terms of accuracy.
Finally, each \lcm is trained three times with different seeds on the weights initialization and the train/test-split.
We average the model predictions across the seeds for all evaluation results reported.

\textbf{Traditional Baselines.}
As a traditional model, we use PostgreSQL's cost model in this study\footnote{While other even more sophisticated traditional models in commercial DBMS such as Microsoft SQL Server exist, we already see as a result of this study that \lcms cannot (yet) improve over PostgreSQL}.
Here, we included two different versions (10.23 and 16.4) to allow a broader comparison and also analyze how traditional cost models evolved over time
In PostgreSQL, the cost of an operator is determined by a weighted sum of the number of disk pages accessed and the amount of data processed in memory.
Note that the cost estimates from PostgreSQL do not represent actual execution times.
Still, they are designed to represent the execution cost and can thus be used for all query optimization tasks.
As such, to make the predictions comparable with the \lcms, we scale the logical costs to the actual runtime with a linear regression model and refer to this approach as \postgresx and \postgresxvi, which is similarly used in other papers \cite{yang2023, zibo_liang_dace_2024, hilprecht2022, zhao2022}.

\textbf{\lcm Implementations.}
For all \lcms, we relied on published source code, which we refer to in our repository\textsuperscript{\ref{code_link}}. 
However, we needed to re-implement some details, such as their ability to work with the same training datasets. 
For instance, \queryformer was hard-coded to work with IMDB only, as it assumed a fixed set of tables and filters. 
We addressed this by first standardizing the various inputs of \lcms (cf. \Cref{tab:taxonomy}), like physical plans, database statistics and sample bitmaps. 
In addition, we maintain feature statistics for each dataset that help to normalize the inputs and to create one-hot encodings (e.g., to encode table or column names) that are required for some \lcms.
We further unified the \lcm training and evaluation pipelines to collect consistent metrics for all models.
However, it is important that all our changes did not change the internal behavior of these models.
\subsection{The Need for New Metrics}
Most works focus on the prediction accuracy over an unseen test dataset by typically reporting the median Q-Error to evaluate how well \lcms predict the execution costs. Often, the median as well as percentiles are reported.
It is defined as follows:
\begin{definition}
\textit{Q-Error ($Q_{50}$)}: The Q-Error is defined as the relative, maximal ratio of an observed label $y$ and its prediction $\hat{y}$ with $Q = \max \left( \hat{y}/{y}, y/\hat{y} \right)$ where $Q=1$ indicates a perfect prediction. 
\end{definition}

However, we argue that this strategy is \textit{not sufficient} to evaluate the applicability of \lcms in query optimization due to two reasons:

\begin{enumerate}[leftmargin=*, nosep]
\item \textbf{Focus on Single Plan Candidates.}
The traditional strategy typically only evaluates \textit{one plan per query} in a workload.
However, the typical task in query optimization is to select one plan out of multiple candidates.
As such, for a study that aims to understand the quality of plan choices, an evaluation methodology needs to enumerate multiple plans for the same query and report metrics that show us how well \lcms can pick the best plan (see next). 

\item \textbf{Focus on Accuracy as only Metric.}
Traditional strategies focus mainly on the overall prediction accuracy of runtimes.
However, the ability of the model to \textit{select the right plan} and to \textit{rank the plan candidates} is much more critical.
Thus, we introduce novel metrics for the corresponding tasks to evaluate the ranking and selection properties of \lcms in the later sections. 
\end{enumerate}

\section{Task 1: Join Ordering} \label{sec:join_order}

Optimizing join order is a crucial task in query optimization as it improves query performance, especially in complex queries involving joins over multiple tables.
In the following, we analyze \emph{how reliably current \lcms reason about join orders} by first describing the detailed experimental setting and additional metrics before presenting the results of several experiments.
\subsection{Evaluation Setup}\label{subsec:join_order_setup}
\noindent\textbf{Experimental Setting.}
Unlike previous evaluation strategies that evaluate the prediction accuracy on a single plan candidate, we compare how well \lcms can be used to pick a join order.
For this, we exhaustively generate \textit{all} possible join permutations for a given workload and let all \lcms predict their execution costs (as a traditional query optimizer would do).
We decided to use exhaustive enumeration to separate concerns and avoid a bad plan being picked based on the enumeration strategy (e.g., by only enumerating left-deep plans).
That way, we analyze how well \lcms provide costs (i.e., runtime) that enable an optimizer to pick good join orders, not how well the enumeration strategy works.
As a workload in this study, we use the JOB-light benchmark \cite{leis_how_2015, kipf2019}, which operates on the IMDB dataset consisting of 70 SPJA-Queries, as it is specifically designed to evaluate the task of join ordering.
In contrast to other datasets like TPC-H, this dataset contains diverse correlations and non-uniform data distributions, increasing the difficulty of the task.

\noindent\textbf{Experimental Metrics.}
As explained before, we define new metrics to evaluate \lcms for the join ordering task as follows:

\begin{enumerate}[leftmargin=*, nosep]
\item To evaluate how \lcms affect the outcome of query optimization in terms of runtime, we introduce the \textit{selected runtime}:
\begin{definition} \label{def:selected_runtime}
\textit{Selected Runtime ($r$)}:
Given a set of plan candidates, the selected runtime, $r$, is defined as the actual runtime $y$ of the plan that the \lcm would choose, i.e., where the prediction $\hat{y}$ is minimal.
Formally, it is:
$r = y_{\arg\min_{i} \hat{y}}$
\end{definition}

\item To report how well \lcms are able to select the optimal out of multiple plans, we introduce \textit{surpassed plans}.
\begin{definition}
\textit{Surpassed Plans ($s_p$)}:
In relation to a selected plan with actual runtime $y$, surpassed plans refer to the proportion of query plans that actually have a longer actual runtime.
$s_p=100\%$ therefore means that the optimal plan was selected, while $s_p=0\%$ stands for the worst choice.
Formally, it is defined as: $s_p = 100 \% * (1/n) * \sum_{i=1}^{n} \text{bool}(y_i > r)$ with total $n$ plans with $s_p \in [0\%, 100\%]$
\end{definition}
    
\item Moreover, it's crucial for \lcms to order query plans by their costs, as the optimizer needs to select from different candidates.
Thus, we report the correlation of actual and predicted runtimes of different query plans with the \textit{rank correlation} that asses the ranking ability of the \lcm towards the query plans \cite{spearman1904}.
\begin{definition}
\textit{Rank Correlation ($\rho$)}:
For our study, we use Spearman's Correlation which is given by: $\rho = 1 - \frac{6 \sum_{i=1}^{n} d_i^2}{n(n^2 - 1)}$, with $\rho \in [0\%, 100\%]$, where $d_i$ is the difference between the ranks of corresponding runtimes, and $n$ is the number of plans.
\end{definition}
    
\item Finally, to evaluate the likelihood of \lcms to pick a non-optimal plan, we report under- and overestimation.
For instance, if the span of \lcm  under- and overestimation is high, it is more likely to select a plan that is, in fact, sub-optimal.
\begin{definition} \textit{Maximal Relative Under- /Overestimation($m_u, m_o$)}:
These metrics indicate by which factor a query plan candidate is over- or underestimated in the worst case. 
For a set of predictions $\hat{y}$ and labels $y$, they are defined as $m_u = min_{i} (y_i / \hat{y}_i)$ and  $m_o = max_{i} (\hat{y}_i / y_i)$.
\end{definition}
\end{enumerate}

\noindent\textbf{Metric Discussion:}
As we show for all experiments, the focus on accuracy alone is not sufficient to evaluate cost models for query optimization, and therefore, we propose various other metrics. 
However, it is important to note that each metric has limitations in what it can show, and only the combination of multiple metrics can help to understand the overall behavior.
For example, for join ordering, correlation metrics are good in helping to understand how well cost models rank plans. 
Still, they fail to analyze the severity of the effect if plans are not ranked well, which can lead to sub-optimal plan selections. 
For this reason, its is crucial to look at the percentage of surpassed plans and the total selected runtime.

\subsection{Example Query \& Metrics}\label{subsec:join_order_anecdotal}
\begin{figure*}[]
    \centering
    \includegraphics[width=\linewidth]{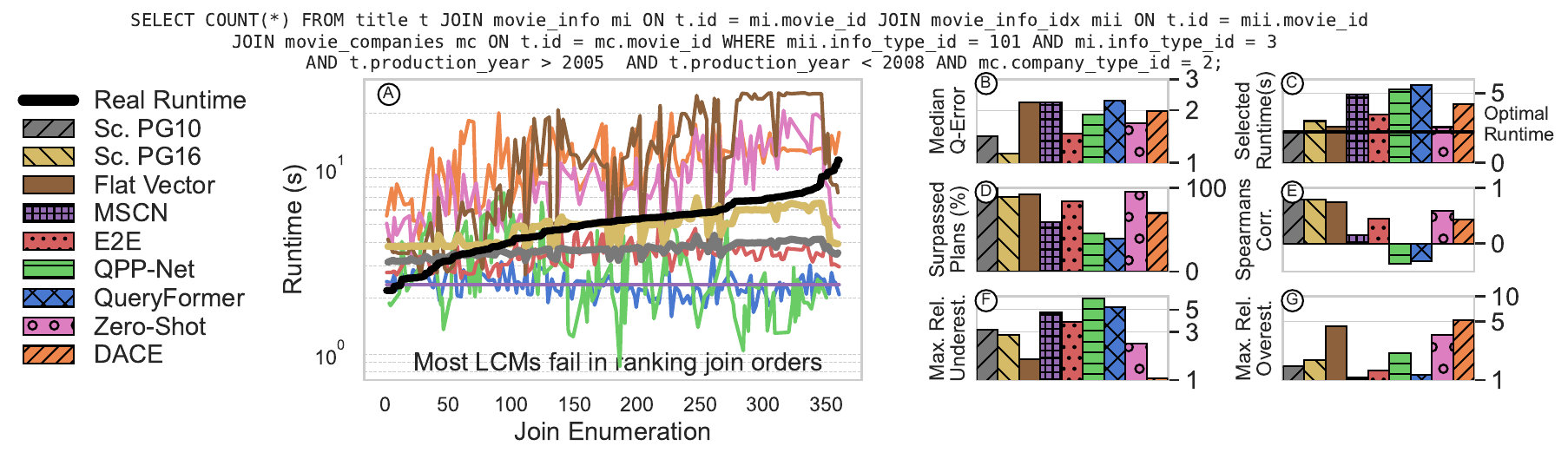}
    \caption{Example Query Nr. 33 from JOB-Light for join ordering. 
    We report model predictions (\circles{A}), overall accuracy (\circles{B}), outcome on query optimization (\circles{C}),  model optimality (\circles{D}), model ranking (\circles{E}) as well as under- and overestimation (\circles{E} \& \circles{F}).
    In this query, \postgresx picks the optimal plan and outperforms all \lcms for most of the metrics.}
    \label{fig:join_order_example}
\end{figure*}

We first report the results of an example query (Query $33$ from JOB-Light) in \Cref{fig:join_order_example} (at the top), as its results are representative of the full study that we conduct in \Cref{sec:join_order_full} and to illustrate our metrics introduced above.
As this query has four tables, there are 120 possible join permutations.
We discuss the results of each subplot (\circles{A} - \circles{G}) in the following:

\noindent\textbf{\circles{A} Model Predictions.}
We sort the plans with the different join permutations by their actual runtime, as presented by the black curve. 
Intuitively, the leftmost plan is optimal, with the shortest runtime of 2.20s, while the worst plan requires 11.17s.
Moreover, we show the predictions of all \lcms in the same plot to get a qualitative impression of their predictions. 
On the first view, the predictions of \lcms often show a behavior where predicted costs do not grow regularly with the actual costs.
In general, this is undesirable behavior, as it reinforces the choice of unfavorable plans since undesirable, false minima can occur in the cost predictions. 
In contrast, both \postgresx (gray curve) and \postgresxvi (gold curve) show that the predicted cost grows with the actual cost.

\noindent\textbf{\circles{B} Median Q-Error.}
To assess the overall prediction accuracy, we report the Q-Error ($Q_{50}$) of the predictions over all join permutations.
Interestingly, for this particular query, \postgresxvi shows the best results with $Q_{50}$=1.23, followed by \postgresx with  $Q_{50}$=1.42.
Moreover, another interesting observation is when looking at \mscn.
While the Q-error is not that bad (with $Q_{50} = 2.23$), the predictions are not helpful for an optimizer at all, as they predict the very same runtime value for all plans. 
Due to its SQL-based featurization, \mscn cannot reason about the join order of the query plan or the query operators and thus, it cannot be used for join ordering or physical plan selection.
Still, we included \mscn, which is often used as a naive baseline for cost predictions \cite{liu2022, hilprecht2022, zhao2022}.

\textbf{\circles{C} Selected Runtime.}
To evaluate how much \lcms affect query optimization, we report the selected runtime ($r$), as this reflects the query runtime of the chosen plan when relying on the cost estimates of a given \lcm.
Here, the best cost model is \postgresx, where a query runtime of $r$=2.20s is achieved.
In contrast, the worst case is when using \queryformer; an optimizer would lead to a plan with a much longer runtime of $r$=5.56s, which is more than twice as optimal.
All other \lcms end up between \postgresx and \queryformer; i.e., no \lcm provides a better plan choice than \postgresx.
Interestingly, \postgresxvi selects a slightly worse plan than \postgresx, which is, however, still near-optimal. 

\textbf{\circles{D} Surpassed Plans.}
Next, we evaluate the surpassed plans ($s_p$).
This shows the fraction of plans outperformed by the selected plan from a given \lcm and thus shows the relative rank of the plan a model picks; i.e., if a model picks a plan on rank 5 of 10 plans, it surpasses $5/10=50\%$ of the plans.
Consistent with the previous results, \postgres beats the other models as it selects the optimal plan in this example ($s_p$=100$\%$).
In contrast, \queryformer (worst \lcm) only surpasses $s_p$=39\% of the plan candidates.

\textbf{\circles{E} Rank Correlation.}
As \lcms crucially need to determine the \textit{rank} of query plans by their costs, it is important to analyze the rank faithfulness of a cost model\footnote{
This observation led to \textit{ranking-based cost models} as an interesting alternative \cite{behr2023, chen2023, zhu2023}.}.
We report the ranking correlation ($\rho$) between the actual and predicted runtimes to evaluate how well the models perform in the ranking task.
In the given example query, both \postgresx and \postgresxvi have the best correlation ($\rho=0.79/ \rho=0.72$) indicating a successful ranking.
In contrast, all \lcm competitors are worse, down to $\rho = -0.23$ for \queryformer. 
Critically, the negative value means that as the actual runtime increases, the predicted runtime tends to decrease, indicating a failure of the ranking task.
This shows that cost predictions of traditional models provide a better ranking of query plans than \lcms for this query.

\textbf{\circles{F} \& \circles{G} Under- and Overestimation.}
We now report underestimation and overestimation of \lcms.
Overall, underestimation and overestimation indicate the probability of a query optimizer not picking the optimal plan.
For example, when significantly underestimating the runtime, a plan might be chosen that is, in fact, a long-running plan.
When looking at \circles{F} \& \circles{G}, we make two interesting observations:
(1) Under- \& overestimation is generally larger with \lcms than for \postgresx and \postgresxvi.
(2) Generally, PostgreSQL tends towards systematic underestimation of execution costs, which similarly has been shown by \cite{leis_how_2015}.
A typical reason for this is the assumption of filter attribute independence, leading to underestimates. 
However, \lcms are prone towards both under- and overestimation.
Overall, this wider range of over- and underestimation of \lcms, in fact, increases the likelihood of picking a non-optimal plan.

\subsection{Full Results on Join Order}\label{sec:join_order_full}
\begin{figure*}
    \centering
\includegraphics[width=\linewidth]{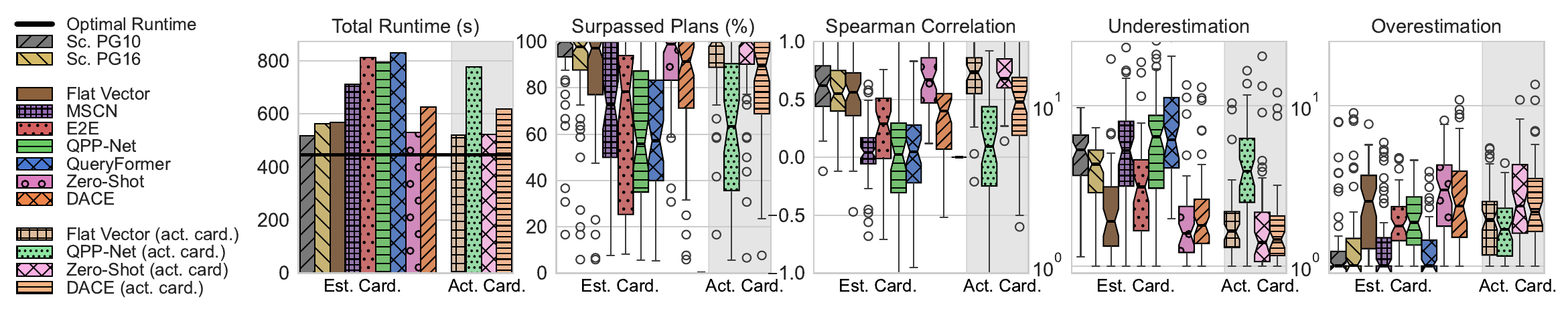}
    \caption{Full Join Ordering Results on JOB-Light benchmark. 
    Traditional models often still outperform \lcms in terms of join ordering across all metrics.
    Using actual cardinalities helps to improve the results of \lcms significantly.}
    \label{fig:join_order_act_cards}
\end{figure*}

To report more representative results, we aggregate the previously discussed metrics over the full JOB-Light benchmark. 
The results can be seen in \Cref{fig:join_order_act_cards}.
Most interestingly, when looking at the total runtime of selected plans, we see that \postgresx still provides the shortest total selected runtime of $r=518s$, followed by \zeroshot with $r=530s$.
The optimal runtime for the workload is at $r$=446s, which assumes a perfect plan selection for every query. 
The worst model again is \queryformer, where the selected runtime for all queries is $r=830s$.
When looking at surpassed plans, \postgresx and \postgresxvi also outperform their \lcm competitors.
Regarding the ranking of plans, as shown by ranking correlation $\rho$, \zeroshot as \lcm actually returns the best results, outperforming traditional approaches.
This is surprising, as \zeroshot still does not lead to plans that outperform them in total runtime.
However, when looking at under- and overestimation, we can see that in line with the anecdotal result, \zeroshot tends to both under- and overestimate at the same time.
This leads to worse plan selections.
Interestingly, the best three performing \lcms are all \textit{database-agnostic} and \flatvector shows a good (often third place) performance, although it uses a simplistic model structure.

\subsection{Impact of Improved Cardinalities}
The optimal order of joins is impacted significantly by intermediate cardinalities, which in turn influence query costs. 
So far, many of the proposed \lcms use \textit{estimated cardinalities} as input derived from PostgreSQL to provide cost estimates for a plan. 
However, the PostgreSQL cardinality estimator frequently produces inaccuracies, sometimes deviating by several orders of magnitude. 
It was shown that better cardinality estimates substantially improve the cost estimation results \cite{leis_how_2015}. 
This raises the question of their impact on cost estimates of \lcms to make better optimizer decisions. 
To investigate this, we train and evaluate all \lcms, that utilize cardinalities as input features (i.e., \flatvector, \qppnet, \zeroshot, \dace, cf. \Cref{tab:taxonomy}), with actual, observed cardinalities rather than estimated ones.
This aims to isolate the effects of cost estimation from that of cardinality estimation.
We repeat the same evaluation from the previous experiment from \Cref{sec:join_order_full} with actual cardinalities instead and report the results in \Cref{fig:join_order_act_cards}, where the new model variants are represented by the lighter-colored bars.
It can be seen that, indeed, perfect cardinalities have a positive impact on the overall results.
For instance, the total runtime of \qppnet decreases from $r$=765s to $r$=644s.
Still, the best \lcm variant, which is \flatvector here, is outperformed by \postgresx, which still relies on estimated cardinalities.
However, as a promising result, we would like to highlight ranking correlation and the underestimation, where adding perfect cardinalities significantly improves all \lcm variants.

\subsection{Accurate Cardinalities for PostgreSQL}\label{subsec:optimizer_comparison}
\begin{figure}
    \centering
    \includegraphics[width=0.8\linewidth]{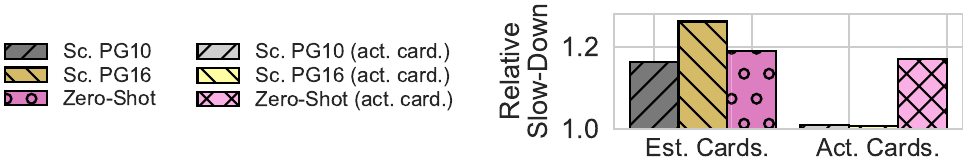}
    \caption{Relative runtime slow-down vs. the optimal execution. PostgreSQL models are closer to the optimum than \lcms, especially when using perfect cardinalities}
    \label{fig:optimizer_runtime}
\end{figure}

Finally, we analyze how optimal the selected plans of \lcms are in relation to the total runtime when providing accurate cardinality estimates for classical cost models. 
We compare \postgresx, \postgresxvi and the best performing \lcm, from the previous experiment (i.e., \zeroshot) and report the relative slow-down of runtime that they achieve compared to the optimal workload runtime for the JOB-Light benchmark.
The results are shown in \Cref{fig:optimizer_runtime} where we see that overall, the execution is 18.9\% slower than the optimum when using \zeroshot.
In contrast, \postgresx achieves a slow-down of $16.23\%$ and \postgresxvi $24.9\%$. 
Importantly, when providing actual cardinalities to the PostgreSQL models, their performance improves drastically towards near-optimal plan selections (1.0\% and 0.8\% slow-down).
In contrast, \zeroshot only gets slightly better, indicating that it cannot yet make full use of information provided by cardinality estimates; i.e. \zeroshot still over- and underestimates costs for individual plans, leading to non-optimal plan selections.

\subsection{Summary \& Takeaways}
The analysis of join ordering indicates that the traditional model, \postgres, continues to outperform \lcms in selecting plans with low runtimes. 
In fact, only in terms of the ability of \textit{ranking} of plans, the traditional models are outperformed by \zeroshot, which demonstrates that \lcms have the ability to improve query optimization.
However, traditional models are not without flaws, as they also still fail to identify faster plans in some cases.
For this task, the three best performing \lcms are all DB-agnostic, and the simple model \flatvector performed relatively well.
Interestingly, \lcms tend to be more precise in predicting costs for query plans close to the optimal plan, while they have more significant prediction errors (over-and under-estimates) for less optimal plans with higher actual runtimes. 
One reason is that training data typically contains more optimal than sub-optimal plans, which comes from the biased training strategy that we discuss in \Cref{sec:lessons}.
As such, in the future, we suggest that \lcms should focus on mitigating over-and under-estimates for such plans by systematically including signals from these plans in the training of \lcms.
Another direction is to extend \lcms to predict not only runtime but also their confidence.

\section{Task 2: Access Path Selection} \label{sec:access_path_selection}
As a second task in our study, we look at access path selection and analyze how much existing \lcms improve over classical models on this task.
An access path determines \textit{how} a database engine retrieves the requested data, with standard methods being sequential scans or index accesses. 
A key issue is that using indexes is not always faster; for small tables or queries that return a large portion of the table, the overhead of accessing the index can outweigh the benefits, making a sequential scan more efficient. 
This makes the selection of access paths a crucial task in query optimization due to its direct impact on database performance \cite{selinger1979}. 
Selecting an access path is typically done by estimating the cost and determining the path with the lowest cost.
In the following, we study whether \lcms improve over classical models by a set of experiments. 
Like before, we first report anecdotal results from individual cases and then provide an analysis with a broader set of workloads.

\subsection{Evaluation Setup}
\noindent\textbf{Experimental Setting.}
In our study, we look at the ability of \lcms to select between sequential scan and index scan using B+-trees that are both supported in PostgreSQL (internally denoted as \texttt{SeqScan}, and \texttt{IndexScan}/\texttt{IndexOnlyScan}). 
There are also other access options, such as bitmap index scans or hash scans, but these are out of the scope of our study.
We will show that even the selection between these two access paths is hard to solve for all models.
Note that the training data contains indexes on the primary keys (PKs) of \textit{all} tables, which is a common setup in databases.

\noindent\textbf{Experimental Metrics.}
For this task, in addition to the previous metrics, we introduce a new metric to describe how accurately a \lcm selects the access path when one class in the decision-making is under- or over-represented (which is often the case for access path selection as we will see in our evaluation).
\begin{definition}
\textit{Balanced Accuracy ($B$)}: Balanced accuracy $B$ assesses the performance of classification models on imbalanced outcomes as it ensures equal performance consideration for both classes.
It represents the arithmetic mean of true positive rate (TPR) and true negative rate (TNR) as follows:
$B = \frac{1}{2} \left( \text{TPR} + \text{TNR} \right)$ with $\text{TPR} = {TP}/(TP + FN)$ and $\text{TNR} = {TN}/(TN + FP)$ and $B \in [0, 1]$.
\end{definition}

\subsection{Example Query \& Metrics} \label{subsec:access_path_anecdotal}
\begin{figure}
    \centering
    \includegraphics[width=0.8\linewidth]{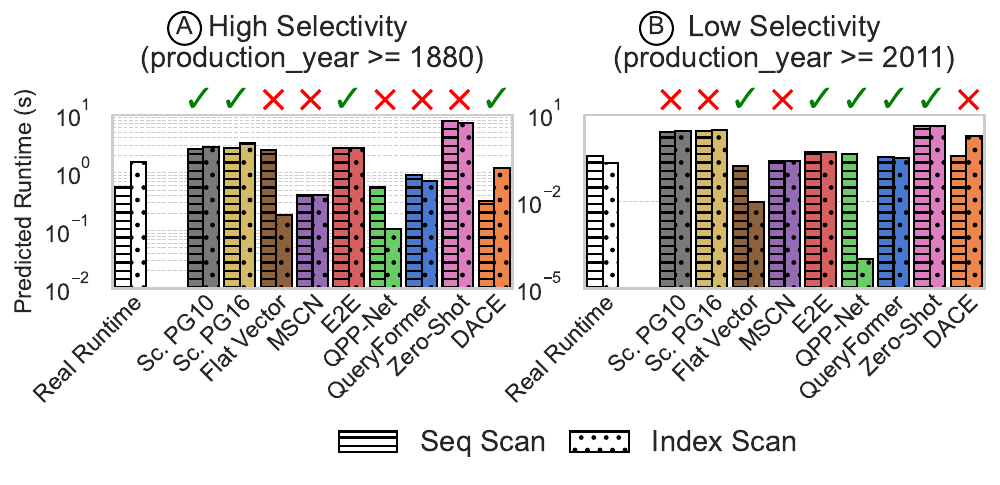}
    \caption{Predictions for table scans on \texttt{movie.production\_year}.
    We show the real runtimes (white bars) against \lcm predictions and indicate correct (\checkmark) and incorrect (\text{\sffamily X}) selections for a high selectivity \circles{A} and low selectivity \circles{B} scenario.}
    \label{fig:scan_cost_anecdote}
\end{figure}

To get an intuition about the access path selection of \lcms, we first analyze how \lcms predict execution costs for either high or low selectivity.
In the following, we will first present an intuitive example to demonstrate our approach and the new metrics before we later present a broader study across various columns.
We select a query that filters on the column \texttt{production\_year} of the table \texttt{title} of the IMDB dataset.
This column contains the production year of different movies and ranges from 1880 to 2019.
Note that the data distribution is skewed, as more movies were produced over a certain period of years.
We perform two corresponding base table accesses on \texttt{production\_year}, each with either \texttt{IndexScan} and \texttt{SeqScan}.
To achieve a low selectivity, we use the attribute \texttt{>=1880} (which basically selects the whole table), and for high selectivity, we use \texttt{>=2011}, which contains just $\approx 20\%$ of the entries.
We report the results in \Cref{fig:scan_cost_anecdote}, where we mark correct (\checkmark) and incorrect (\text{\sffamily X}) access path selections based on the \lcm predictions.
Moreover, we present the actual runtimes and the model predictions for both scenarios.
For the low selectivity query \circles{A}, we can see, based on the actual runtime (left-most two bars), that the sequential scan is indeed faster (0.71s) than the index scan (1.78s). 
In contrast, the high selectivity query \circles{B} is the reverse.
As we can also see, the cost predictions of the \lcms is often the opposite and lead to wrong access path selection.
In fact, five of nine models (\flatvector, \mscn, \zeroshot, \queryformer, \qppnet) select the 2.5$\times$ slower index scan for the high selectivity (shown in \circles{A}).
In contrast, for the low selectivity query (shown in \circles{B}), the sequential scan is actually  5$\times$ slower (0.35s) than the index scan (0.07s).
Only two \lcms (\mscn, \dace) select the wrong access pass. 
However, if we look at the cost estimates, we can see that estimates are for many \lcms not very indicative since the estimated runtime for scan vs. index are very close, which means that access path selection is highly unstable for this query. 
Overall, this indicates that \lcms are struggling with the correct access path selection and their accuracy changes over the selectivity.

\subsection{Access Path Selection over Selectivities}
\label{subsec:access_path_selectivity}
\begin{figure*}
    \centering
    \includegraphics[width=\linewidth]{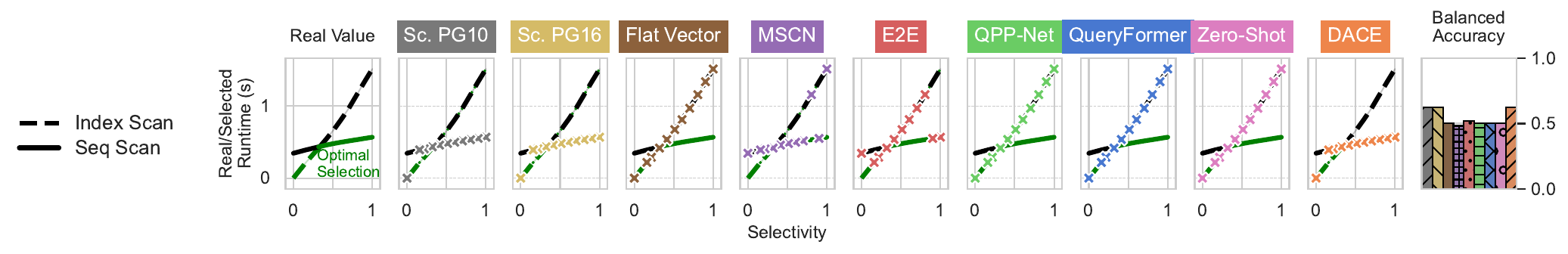}
    \caption{Real scan costs and selected runtimes over different selectivities for scanning the column \texttt{title.production\_year} of IMDB dataset with either sequential or index scan. 
    Many \lcms would select the \texttt{IndexScan} regardless of the selectivity.}
    \label{fig:scan_costs_over_col}
\end{figure*}

Next, we show the results on a broader set of selectivities over different predicates. 
To achieve this, we generate appropriate filter literals for the same query as in the previous section.
More precisely, we scan all movies by \texttt{production\_year} with the \texttt{>=} operator and vary the predicates to ensure equal steps in the domain of selectivities.
We analyze the runtime for each query by selecting the access path according to the cost estimates provided by the \lcm.
The results are shown in \Cref{fig:scan_costs_over_col}.
The real runtimes of the access methods over the selectivity for the sequential and index scans are reported on the very left.
In green, we show the runtime if a cost model selects the optimal path.
As expected, for small selectivities, it is beneficial to select an index scan (dashed line).
When the selectivity gets larger than 0.3, the sequential scan (solid line) is a better choice.
Next, we show in \Cref{fig:scan_costs_over_col} which access paths PostgreSQL selects (second \& third plot)  and contrast this with the access path selection of \lcms (fourth to last plot).
The selected access paths of the cost models are shown by a cross per query we executed.
When looking at the model selections, we see that actually \textit{no single model} always selects the correct access method.
In fact, many models always select the same access method regardless of the selectivity; more details later.
For instance, \flatvector, \zeroshot, \qppnet and \queryformer always select the index scan while other models select different access paths depending on the selectivity.
However, other models like \mscn and \etoe show a highly unstable behavior, especially for higher selectivities, where they randomly switch between access paths.
While \postgresx and \postgresxvi show the desired behavior, they also mispredict access paths for some queries having lower selectivities. 
This behavior is followed by \dace, which learns from PostgreSQL costs.

To aggregate results over the access path choices, we report the balanced prediction accuracy $B$ for each model in \Cref{fig:scan_costs_over_col} (right).
As analyzed before, both PostgreSQL models and \dace achieve the best overall balanced accuracy that reports classification performance with $B=0.62$ and outperform the remaining \lcms.
Overall, these numbers are, however, still not satisfying, as they indicate that all cost models (even classical ones) still struggle to select the best access method.

\begin{table}[]
\resizebox{0.8\linewidth}{!}{
\begin{tabular}{l|c|c|cc|cc|cc}
\multirow{2}{*}{\textbf{Dataset}} & \multirow{2}{*}{\textbf{\begin{tabular}[c]{@{}c@{}}Number \\ of Tables\end{tabular}}} & \multirow{2}{*}{\textbf{\begin{tabular}[c]{@{}c@{}}AVG NaN \\ ratio (\%)\end{tabular}}} & \multicolumn{2}{c|}{\textbf{\begin{tabular}[c]{@{}c@{}}Columns \\ per Table (\#)\end{tabular}}} & \multicolumn{2}{c|}{\textbf{\begin{tabular}[c]{@{}c@{}}Table \\ Length (\#)\end{tabular}}} & \multicolumn{2}{c}{\textbf{\begin{tabular}[c]{@{}c@{}}Distinct\\ Column Values (\#)\end{tabular}}} \\
 &  &  & \multicolumn{1}{c|}{min} & max & \multicolumn{1}{c|}{min} & max & \multicolumn{1}{c|}{min} & max \\ \hline
Baseball & 25 & 9.70 & \multicolumn{1}{c|}{25} & 48 & \multicolumn{1}{c|}{520} & $1.38\times10^6$ & \multicolumn{1}{c|}{2} & $1.64\times10^4$ \\ \hline
IMDB & 15 & 20.91 & \multicolumn{1}{c|}{4} & 49 & \multicolumn{1}{c|}{4} & $1.48\times10^7$ & \multicolumn{1}{c|}{2} & $3.62\times10^7$ \\ \hline
TPC-H & 8 & 0.00 & \multicolumn{1}{c|}{2} & 12 & \multicolumn{1}{c|}{5} & $1.50\times10^6$ & \multicolumn{1}{c|}{2} & $1.50\times10^6$ \\
\hline
\end{tabular}
}
\caption{Statistics of the datasets Baseball, IMDB and TPC-H used to evaluate access path selection of \lcms}
\label{tab:dataset_statistics}
\end{table}

\subsection{Access Path Selection across Queries}
Here, we show the results of a broader study, which confirms similar issues as observed in the previous experiment when selecting access paths for tables and columns with different sizes and data characteristics. 
In this evaluation, we use three different datasets (Baseball, IMDB and TPC-H) from \cite{hilprecht2022}, as they differ in their data characteristics by various dimensions, as shown in \Cref{tab:dataset_statistics}.
For this experiment, we focus on columns that do not have an index according to the training data (i.e., PK columns) and ask \lcms about access path costs if an index were available. 
Moreover, we exclude columns that have more than 70\% of missing values as this leads to large cardinality errors, which we want to isolate from this study.
Finally, we focus on numeric data types to obtain reliable percentiles, as they allow range filter predicates to vary the selectivities precisely.
The overall results are in \Cref{tab:scan_costs_over_datasets}, where we report average accuracy $B$ over all columns (last column in gray).
As seen previously, \postgresx, \postgresxvi, and \dace again perform best with $B\approx 0.64$ on average across all columns and tables in this broader study.
Overall, the simple \flatvector model performs second best.
Although some \lcms (e.g., \qppnet) learn specifically from statistics and histograms, they do not show better performance.
It is, therefore, questionable whether \lcms can derive meaningful information from these artifacts.
Another interesting observation is that results across columns vary for \lcms, ranging from low to high accuracy (marked in bold) across different columns.
Moreover, some \lcms are not better than randomly guessing the access path, which would result in an accuracy of $0.5$, e.g., \queryformer and \qppnet.
Overall, this is far from satisfactory for robustly solving the task of access path selection.


\begin{table*}[]
\resizebox{1\linewidth}{!}{
\begin{tabular}{l|ccccccc|cccccc|cccccccc|
>{\columncolor[HTML]{EFEFEF}}c |}
 & \multicolumn{7}{c|}{\textbf{Baseball}} & \multicolumn{6}{c|}{\textbf{IMDB}} & \multicolumn{8}{c|}{\textbf{TPC-H}} & \cellcolor[HTML]{EFEFEF} \\ \cline{2-22}
\multirow{-2}{*}{\textbf{\diagbox{\lcm}{Tab./Col.}}} & \multicolumn{1}{c|}{\rotatebox{90}{\parbox{2cm}{batting\\AB}}} & \multicolumn{1}{c|}{\rotatebox{90}{\parbox{2cm}{batting\\G\_batting}}} & \multicolumn{1}{c|}{\rotatebox{90}{\parbox{2cm}{halloffame\\needed}}} & \multicolumn{1}{c|}{\rotatebox{90}{\parbox{2cm}{managers\\L}}} & \multicolumn{1}{c|}{\rotatebox{90}{\parbox{2cm}{managers\\W}}} & \multicolumn{1}{c|}{\rotatebox{90}{\parbox{2cm}{mgrhalf\\L}}} & \rotatebox{90}{\parbox{2cm}{mgrhalf\\W}} & \multicolumn{1}{c|}{\rotatebox{90}{\parbox{2cm}{aka\_name\\person\_id}}} & \multicolumn{1}{c|}{\rotatebox{90}{\parbox{2cm}{cast\_info\\movie\_id}}} & \multicolumn{1}{c|}{\rotatebox{90}{\parbox{2cm}{cast\_info\\nr\_order}}} & \multicolumn{1}{c|}{\rotatebox{90}{\parbox{2cm}{person\_inf\\person\_id}}} & \multicolumn{1}{c|}{\rotatebox{90}{\parbox{2cm}{title\\episode\_nr}}} & \rotatebox{90}{\parbox{2cm}{title\\prod\_year}} & \multicolumn{1}{c|}{\rotatebox{90}{\parbox{2cm}{lineitem\\l\_ext.price}}} & \multicolumn{1}{c|}{\rotatebox{90}{\parbox{2cm}{lineitem\\l\_partkey}}} & \multicolumn{1}{c|}{\rotatebox{90}{\parbox{2cm}{part\\p\_retailprice}}} & \multicolumn{1}{c|}{\rotatebox{90}{\parbox{2cm}{part\\p\_size}}} & \multicolumn{1}{c|}{\rotatebox{90}{\parbox{2cm}{partsupp\\ps\_availqty}}} & \multicolumn{1}{c|}{\rotatebox{90}{\parbox{2cm}{partsupp\\ps\_partkey}}} & \multicolumn{1}{c|}{\rotatebox{90}{\parbox{2cm}{partsupp\\ps\_suppkey}}} & \rotatebox{90}{\parbox{2cm}{partsupp\\ps\_supplycost}} & \multirow{-2}{*}{\cellcolor[HTML]{EFEFEF}\textbf{\begin{tabular}[c]{@{}c@{}}Total \\ AVG\end{tabular}}} \\ \hline
\multicolumn{1}{|l|}{\textbf{\cellcolor[HTML]{9B9B9B}\textcolor{white}{\postgresx}}} & \multicolumn{1}{c|}{\textbf{0.67}} & \multicolumn{1}{c|}{\textbf{0.67}} & \multicolumn{1}{c|}{0.50} & \multicolumn{1}{c|}{\textbf{0.6}} & \multicolumn{1}{c|}{0.60} & \multicolumn{1}{c|}{0.50} & 0.50 & \multicolumn{1}{c|}{0.62} & \multicolumn{1}{c|}{0.75} & \multicolumn{1}{c|}{0.57} & \multicolumn{1}{c|}{\textbf{0.6}} & \multicolumn{1}{c|}{\textbf{0.58}} & \textbf{0.62} & \multicolumn{1}{c|}{\textbf{0.67}} & \multicolumn{1}{c|}{\textbf{0.75}} & \multicolumn{1}{c|}{0.6} & \multicolumn{1}{c|}{0.5} & \multicolumn{1}{c|}{0.75} & \multicolumn{1}{c|}{\textbf{1.0}} & \multicolumn{1}{c|}{0.67} & 0.67 & \textbf{0.64} \\ \hline
\multicolumn{1}{|l|}{\textbf{\cellcolor[HTML]{d5bb67}\textcolor{white}{\postgresxvi}}} & \multicolumn{1}{c|}{\textbf{0.67}} & \multicolumn{1}{c|}{\textbf{0.67}} & \multicolumn{1}{c|}{0.50} & \multicolumn{1}{c|}{\textbf{0.6}} & \multicolumn{1}{c|}{0.60} & \multicolumn{1}{c|}{0.50} & 0.50 & \multicolumn{1}{c|}{0.62} & \multicolumn{1}{c|}{0.75} & \multicolumn{1}{c|}{0.57} & \multicolumn{1}{c|}{\textbf{0.6}} & \multicolumn{1}{c|}{\textbf{0.58}} & \textbf{0.62} & \multicolumn{1}{c|}{\textbf{0.67}} & \multicolumn{1}{c|}{\textbf{0.75}} & \multicolumn{1}{c|}{0.6} & \multicolumn{1}{c|}{0.5} & \multicolumn{1}{c|}{0.75} & \multicolumn{1}{c|}{0.88} & \multicolumn{1}{c|}{0.67} & 0.67 & \textbf{0.64} \\ \hline
\multicolumn{1}{|l|}{\textbf{\cellcolor[HTML]{8c613c}\textcolor{white}{\texttt{Flat Vector}}}} & \multicolumn{1}{c|}{0.58} & \multicolumn{1}{c|}{0.46} & \multicolumn{1}{c|}{0.50} & \multicolumn{1}{c|}{0.50} & \multicolumn{1}{c|}{0.40} & \multicolumn{1}{c|}{\textbf{0.83}} & \textbf{1.0} & \multicolumn{1}{c|}{0.38} & \multicolumn{1}{c|}{0.50} & \multicolumn{1}{c|}{0.50} & \multicolumn{1}{c|}{0.40} & \multicolumn{1}{c|}{0.33} & 0.5 & \multicolumn{1}{c|}{0.23} & \multicolumn{1}{c|}{0.36} & \multicolumn{1}{c|}{0.4} & \multicolumn{1}{c|}{0.38} & \multicolumn{1}{c|}{\textbf{0.94}} & \multicolumn{1}{c|}{0.71} & \multicolumn{1}{c|}{\textbf{1.0}} & \textbf{1.0} & 0.57 \\ \hline
\multicolumn{1}{|l|}{\textbf{\cellcolor[HTML]{956cb4}\textcolor{white}{\texttt{MSCN}}}} & \multicolumn{1}{c|}{0.46} & \multicolumn{1}{c|}{0.69} & \multicolumn{1}{c|}{0.48} & \multicolumn{1}{c|}{0.30} & \multicolumn{1}{c|}{\textbf{0.73}} & \multicolumn{1}{c|}{0.50} & 0.50 & \multicolumn{1}{c|}{0.61} & \multicolumn{1}{c|}{0.25} & \multicolumn{1}{c|}{0.32} & \multicolumn{1}{c|}{0.33} & \multicolumn{1}{c|}{0.5} & 0.48 & \multicolumn{1}{c|}{0.54} & \multicolumn{1}{c|}{0.39} & \multicolumn{1}{c|}{\textbf{0.72}} & \multicolumn{1}{c|}{0.46} & \multicolumn{1}{c|}{0.39} & \multicolumn{1}{c|}{0.25} & \multicolumn{1}{c|}{0.48} & 0.25 & 0.46 \\ \hline
\multicolumn{1}{|l|}{\textbf{\cellcolor[HTML]{d65f5f}\textcolor{white}{\texttt{End-To-End}}}} & \multicolumn{1}{c|}{0.31} & \multicolumn{1}{c|}{0.50} & \multicolumn{1}{c|}{0.50} & \multicolumn{1}{c|}{0.50} & \multicolumn{1}{c|}{0.50} & \multicolumn{1}{c|}{0.58} & 0.50 & \multicolumn{1}{c|}{\textbf{0.75}} & \multicolumn{1}{c|}{0.25} & \multicolumn{1}{c|}{0.50} & \multicolumn{1}{c|}{0.28} & \multicolumn{1}{c|}{0.5} & 0.52 & \multicolumn{1}{c|}{0.5} & \multicolumn{1}{c|}{0.5} & \multicolumn{1}{c|}{0.4} & \multicolumn{1}{c|}{0.5} & \multicolumn{1}{c|}{0.5} & \multicolumn{1}{c|}{0.5} & \multicolumn{1}{c|}{0.33} & 0.5 & 0.47 \\ \hline
\multicolumn{1}{|l|}{\textbf{\cellcolor[HTML]{6acc64}\textcolor{white}{\texttt{QPP-Net}}}} & \multicolumn{1}{c|}{0.50} & \multicolumn{1}{c|}{0.50} & \multicolumn{1}{c|}{0.50} & \multicolumn{1}{c|}{0.50} & \multicolumn{1}{c|}{0.50} & \multicolumn{1}{c|}{0.50} & 0.50 & \multicolumn{1}{c|}{0.50} & \multicolumn{1}{c|}{\textbf{1.0}} & \multicolumn{1}{c|}{\textbf{0.79}} & \multicolumn{1}{c|}{0.50} & \multicolumn{1}{c|}{0.5} & 0.5 & \multicolumn{1}{c|}{0.5} & \multicolumn{1}{c|}{0.5} & \multicolumn{1}{c|}{0.5} & \multicolumn{1}{c|}{0.5} & \multicolumn{1}{c|}{0.5} & \multicolumn{1}{c|}{0.5} & \multicolumn{1}{c|}{0.5} & 0.5 & 0.54 \\ \hline
\multicolumn{1}{|l|}{\textbf{\cellcolor[HTML]{4878d0}\textcolor{white}{\texttt{QueryFormer}}}} & \multicolumn{1}{c|}{1.0} & \multicolumn{1}{c|}{0.50} & \multicolumn{1}{c|}{0.50} & \multicolumn{1}{c|}{0.50} & \multicolumn{1}{c|}{0.60} & \multicolumn{1}{c|}{0.50} & 0.50 & \multicolumn{1}{c|}{0.38} & \multicolumn{1}{c|}{0.56} & \multicolumn{1}{c|}{0.36} & \multicolumn{1}{c|}{0.50} & \multicolumn{1}{c|}{0.25} & 0.5 & \multicolumn{1}{c|}{0.38} & \multicolumn{1}{c|}{0.22} & \multicolumn{1}{c|}{0.5} & \multicolumn{1}{c|}{0.5} & \multicolumn{1}{c|}{0.5} & \multicolumn{1}{c|}{0.5} & \multicolumn{1}{c|}{0.5} & 0.5 & 0.50 \\ \hline
\multicolumn{1}{|l|}{\textbf{\cellcolor[HTML]{dc7ec0}\textcolor{white}{\texttt{ZeroShot}}}} & \multicolumn{1}{c|}{0.42} & \multicolumn{1}{c|}{0.60} & \multicolumn{1}{c|}{0.50} & \multicolumn{1}{c|}{0.50} & \multicolumn{1}{c|}{0.50} & \multicolumn{1}{c|}{0.50} & 0.50 & \multicolumn{1}{c|}{0.38} & \multicolumn{1}{c|}{0.56} & \multicolumn{1}{c|}{0.36} & \multicolumn{1}{c|}{0.50} & \multicolumn{1}{c|}{0.5} & 0.5 & \multicolumn{1}{c|}{0.5} & \multicolumn{1}{c|}{0.25} & \multicolumn{1}{c|}{0.63} & \multicolumn{1}{c|}{0.36} & \multicolumn{1}{c|}{0.25} & \multicolumn{1}{c|}{0.5} & \multicolumn{1}{c|}{0.33} & 0.46 & 0.46 \\ \hline
\multicolumn{1}{|l|}{\textbf{\cellcolor[HTML]{ee864a}\textcolor{white}{\texttt{DACE}}}} & \multicolumn{1}{c|}{\textbf{0.67}} & \multicolumn{1}{c|}{\textbf{0.67}} & \multicolumn{1}{c|}{\textbf{0.62}} & \multicolumn{1}{c|}{\textbf{0.6}} & \multicolumn{1}{c|}{0.60} & \multicolumn{1}{c|}{0.50} & 0.50 & \multicolumn{1}{c|}{0.62} & \multicolumn{1}{c|}{0.75} & \multicolumn{1}{c|}{0.50} & \multicolumn{1}{c|}{\textbf{0.6}} & \multicolumn{1}{c|}{\textbf{0.58}} & \textbf{0.62} & \multicolumn{1}{c|}{\textbf{0.67}} & \multicolumn{1}{c|}{\textbf{0.75}} & \multicolumn{1}{c|}{\textbf{0.6}} & \multicolumn{1}{c|}{\textbf{0.62}} & \multicolumn{1}{c|}{0.75} & \multicolumn{1}{c|}{0.86} & \multicolumn{1}{c|}{0.67} & 0.67 & \textbf{0.64} \\ \hline
\end{tabular}
}
\caption{Balanced accuracy $B$ of \lcms when selecting access paths for different workloads, tables, and columns.}
\label{tab:scan_costs_over_datasets}
\end{table*}

\subsection{Access Path Preferences} \label{subsec:access_path_preference}
\begin{figure}
    \centering
    \includegraphics[width=0.8\linewidth]{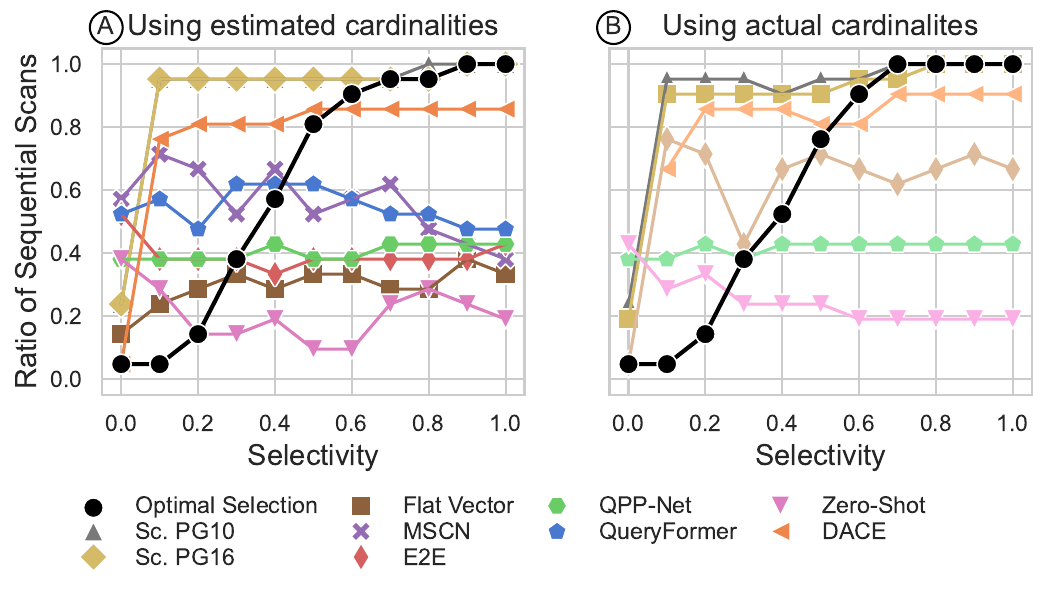}
    \caption{Scan preference of \lcms over selectivity using either estimated \circles{A} or actual cardinalities \circles{B}}
    \label{fig:scan_preference}
\end{figure}

To analyze why many \lcms often choose the wrong access paths, we analyze their selection preferences across all queries.
For this, we use all queries of the previous experiment and show in \Cref{fig:scan_preference}\circles{A} the averaged ratio of selected table scans broken down by selectivities. 
The index scan ratio is reflected in this experiment as a $1 - [\text{ratio of table scans}]$.
In \Cref{fig:scan_preference}\circles{A}, the black line shows the optimal selection.
For the \lcms (colored lines), the observations can be divided into two groups: 
(1) Overall, \postgresx, \postgresxvi and \dace are closest to the optimal selection and follow the trend across the selectivities, i.e., they select more index scans for small selectivities and more table scans for large selectivities.
However, they select sequential scans far too early for too small selectivities.
(2) The other \lcms, on the other hand, seem not to understand the effect of selectivity at all, as the ratio of table scans is almost constant across varying selectivities. 
Moreover, there seems to be some (static) preference towards index scans for many \lcms.
To explain why many \lcms have this bias, we analyzed the training data. 
Surprisingly, it contained most often sequential scans ($\approx90\%$) and a few index scans ($\approx10\%$).
Moreover, index scans were only used when they were really beneficial for queries, and no negative examples were included. 
Therefore, \lcms learn that index scans seem highly promising without understanding the downsides for high selectivities.

\subsection{The Effect of Improved Cardinalities}
Finally, as for the join order task, we want to see the effect of cardinality estimates, which are input to many cost models, and see if more accurate estimates lead to better decisions for cost models.
For this purpose, we repeat the previous experiment with perfect cardinalities to estimate their effect on the access path selection.
Interestingly, as shown in \Cref{fig:scan_preference}\circles{B}, the access path selection improves slightly for some \lcm which now better follow the optimal choice, e.g., \flatvector.
However, other \lcms are still not able to catch the trend.
We analyzed this behavior and found that the cardinality estimates for these simple scan queries are already highly accurate (with a $Q_{50} <= 1.05$). 
Thus, further improving cardinalities does not help this task very much.

\subsection{Summary \& Takeaways} \label{sec:access_path_discussion}
Overall, no single \lcm delivers convincing results in access path selection. 
Different from join ordering, the main reason is still that the costs of access paths are not really understood.
However, incorporating database statistics and sample bitmaps into \lcms was not beneficial.
Interestingly, \lcms are biased in their selection towards index scans across the selectivity range, which indicates that they learned that index scans are always beneficial due to bias in the training data.
To mitigate this bias, \lcms need to learn from execution costs for both access paths across selectivity.
We present strategies and initial results for this in \Cref{sec:lessons}.

\section{Task 3: Physical Operator Selection}\label{sec:physical_operator}
The task of physical operator selection refers to the process of choosing the most efficient algorithm to execute a given query operator. 
This selection is crucial, as it directly impacts the query performance and resource utilization. 
The complexity of this task arises from understanding the combined effects of algorithm complexity and various other factors, such as the data distribution, available indexes, hardware capabilities, and the specific characteristics of the workload. 
In this section, we report how well recent \lcms are able to select physical operators using physical operator selection for joins as an important example.

\subsection{Evaluation Setup}
\noindent\textbf{Experimental Setting.}
For this experiment, we generate various queries that join two tables with join predicates on the foreign key relationships as these are the most common joins, like: \texttt{SELECT COUNT(*) FROM title,movie\_info\_idx WHERE title.id=movie\_info\_idx.movie\_id AND title.production\_year=2009}.
We use \texttt{COUNT}-expressions so as not to deviate too much from the training data, which also makes typical use of them.
We execute each query three times with three different join algorithms to get the true runtimes, i.e. \texttt{Hash Join} (HJ), \texttt{Sort Merge Join}, (SMJ) \texttt{Indexed Nested Loop Join} (INLJ)\footnote{We only used INLJ joins, as primary keys always used an index in our setup.} which are algorithms that PostgreSQL supports.
For all queries, we also obtain the predictions of \lcms.

\noindent\textbf{Experimental Metrics.}
For this study, we introduce a new metric that determines how often the \lcm picks the plan with the optimal physical operator algorithm.
\begin{definition}
\textit{Pick Rate (p)}: The pick rate reports the percentage of $p$ out of $n$ query plans that gives how often a cost model would pick the optimal plan, e.g., where the plan with the minimal prediction has the lowest actual runtime.
\end{definition}
\subsection{Example Query \& Metrics} \label{subsec:physical_operator_anecdotal}
\begin{figure}
    \centering
    \includegraphics[width=0.8\linewidth]{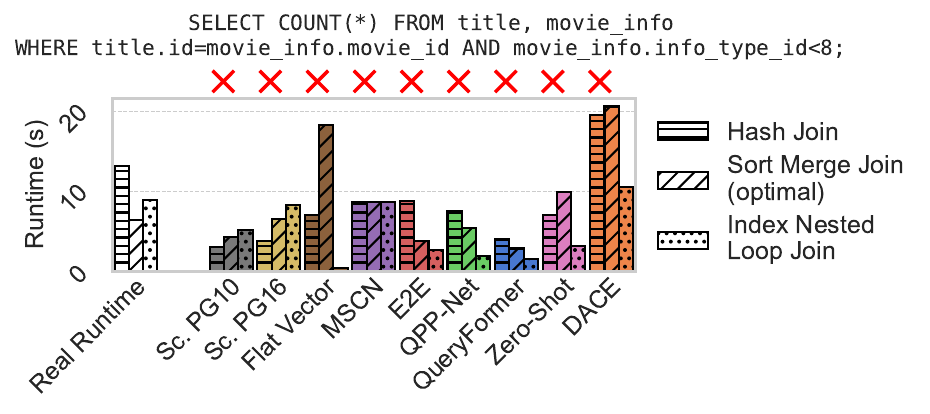}
    \caption{Predictions for physical operators for a two-way join on the IMDB dataset. 
    We show the real runtimes against \lcm predictions and indicate correct (\checkmark) and incorrect (\text{\sffamily X}) selections. 
    No single \lcm would pick the fastest join (SMJ).}
\label{fig:physical_operator_anecdote}
\end{figure}

\noindent To first get an impression of how \lcms select physical operators and how our novel metrics are applied, we illustrate in \Cref{fig:physical_operator_anecdote} the runtimes for a representative example query of the IMDB dataset when using three different physical operators (left) and the model predictions (right).
For the given query, SMJ is the most optimal selection according to the runtime of around $6.47s$, while INLJ and HJ have longer runtimes of up to $13.19s$ (left-most bars).
Interestingly, neither the PostgreSQL cost models nor a single \lcm select the optimal physical operator.
In fact, most models prefer the INLJ, although it has a longer runtime of $8.94s$.
Note that \postgresx and \postgresxvi also fail here, as they select the worst operator (HJ) with $13.19s$.
\subsection{Full Results on Physical Operator Selection}
\label{sec:physical_operator_full}
We now validate these initial findings for a broader set of queries and datasets.
For this, we repeat the previous experiment with each 100 queries on the three different datasets IMDB, Baseball, and TPC-H (as described in \Cref{tab:dataset_statistics}).

\begin{figure}
    \centering
    \includegraphics[width=0.9\linewidth]{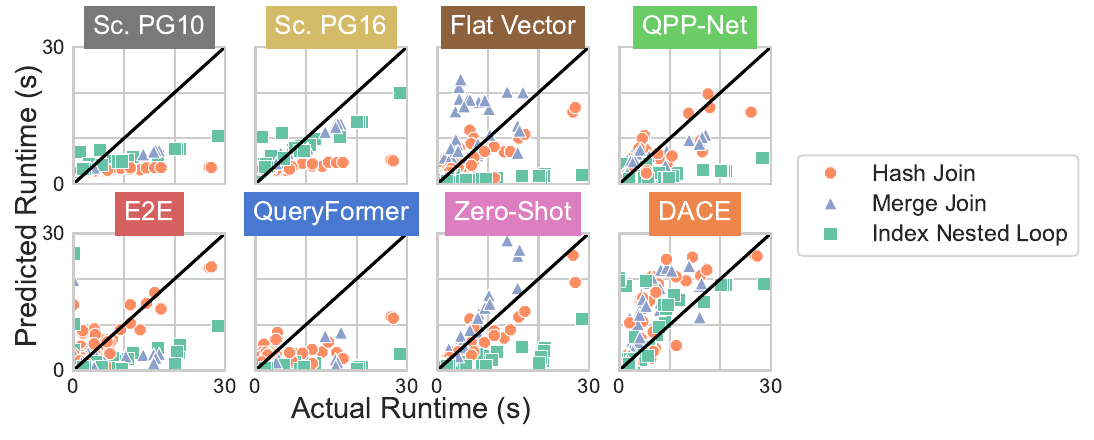}
    \caption{Predicted vs. actual runtimes for different join types on 300 queries (on  IMDB, TPC-H and baseball datasets)}
\label{fig:physical_operator_all}
\end{figure}

\textbf{All Predictions}: 
We visualize the predictions vs. the actual runtimes in \Cref{fig:physical_operator_all}.
As each of the 300 queries has three different plan candidates with different joins, each subplot contains 900 predictions.
Overall, for many \lcms, we observe huge differences between the operator type and the predicted costs.
For example, \zeroshot overall shows good correlations between the actual and predicted runtime for each operator type.
However, while Hash Joins are precisely estimated, Index Nested Loop Joins are systematically underestimated, and Sort-Merge-Joins are overestimated roughly by a constant factor.
Similarly, \postgresx and \postgresxvi show a linear trend between the predicted and actual runtime but overall tend to underestimate.
In contrast, other \lcms like \etoe or \dace show a noisy behavior that does not show any linear function of the operator type.
However, both types of behavior lead to a sub-optimal operator selection.
Interestingly, for some \lcms, we observe a consistent ranking within an operator type (e.g. INLJ for \flatvector) but not across operator types, which makes it particularly hard to select the optimal operator. 

\begin{figure*}
    \centering
    \includegraphics[width=\linewidth]{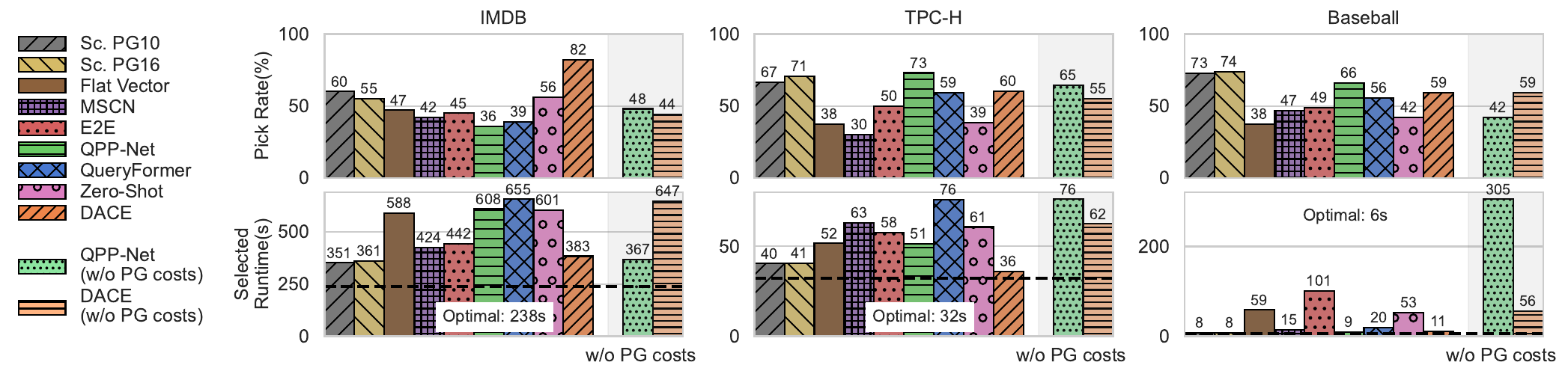}
    \caption{Pick rate and selected runtime for physical operator selection over 100 queries of three datasets (IMDB, TPC-H, Baseball). \lcms are able to achieve a better performance than \postgres. However, the performance deteriorates when removing estimated PostgreSQL costs from the input features of \dace and \qppnet.}
    \label{fig:physical_operator_selection_full}
\end{figure*}

\textbf{Aggregated Results}:
To provide a more quantitative evaluation, we report aggregated results in \Cref{fig:physical_operator_selection_full}.
We report the pick rate $p$ (upper row) for each workload and the selected total runtime $r$ vs. the runtime when optimal operators are chosen (lower row).
Interestingly, when looking at the pick rate on IMDB, \lcms perform comparably well to \postgresx and \postgresxvi, and some even outperform them on IMDB and TPC-H.
For instance, \dace as a database-agnostic model achieves a pick rate of $p=82\%$ on IMDB, whereas \postgresx only achieves $p=60\%$.
A similar trend is observed for the TPC-H dataset.
However, for baseball, \postgresxvi achieves the best pick rate with $r=74\%$.
A slightly different trend is observed when looking at the overall selected runtime.
For IMDB and Baseball, \postgresx and \postgresxvi achieve the lowest runtime, closely followed by \dace. 
On TPC-H, \dace outperforms the other models slightly.
These results show that \lcms already have the potential to be competitive with traditional models on this task while again not providing significant benefits.

\subsection{Learning From PostgreSQL Costs}
\dace and \qppnet use estimated PostgreSQL costs as input features to predict the cost of queries as shown in \Cref{tab:taxonomy}.
They, therefore, function as \textit{hybrid} approaches, as they build on PostgreSQL estimates.
That way, these models utilize the expert knowledge inherent in the PostgreSQL cost model.
In this ablation study, we thus want to explore the contribution of PostgreSQL costs to the estimates.
Thus, we train a variant of these two models \qppnet and \dace without PostgreSQL costs as input, i.e., we removed them from the featurization.
The results can be seen in \Cref{fig:physical_operator_selection_full} (light bars at the right in each plot) 
As expected, the pick rate of \dace on the IMDB dataset decreases from $p=82\%$ to $p=44\%$, and the selected runtime increases from $r=383$ to $r=647s$.
A similar observation also holds on the other datasets and also for \qppnet -- except for one case where surprisingly \qppnet gets better on IMDB.
Thus, it becomes clear that the PostgreSQL costs typically are an important signal and should be included in future \lcms, as we will discuss.
\subsection{Operator Breakdown \& Preferences}
\begin{figure*}
    \centering
    \includegraphics[width=1\linewidth]{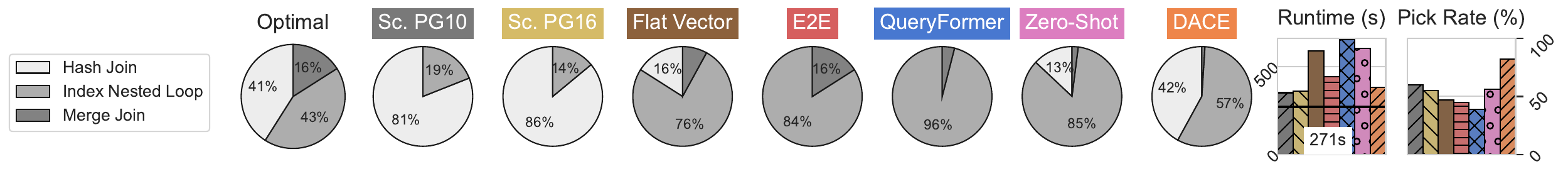}
    \caption{Breakdown of selected operators for physical operator selection on the IMDB test queries.}
    \label{fig:physical_operator_breakdown}
\end{figure*}

In this experiment, we want to understand where \lcms make mistakes when selecting physical operators.
For this, we show the distributions of the selected operator types for the 100 queries on the IMDB dataset compared to the optimal distribution in \Cref{fig:physical_operator_breakdown}.
As shown by the optimal distribution (left), in 43\% of the queries, an INLJ is best, while in 41\%, it is the HJ, and in 16\%, the SMJ.
However, we see fundamentally different operator selections when looking at the selections when using the cost models.
First, \postgresx and \postgresxvi both show a strong tendency to over-select HJs.
Looking at \lcms, we see a very different picture: 
Most \lcms have an over-preference for INLJs.
For example, \queryformer chooses an INLJ even in 95\%, and \zeroshot in 85\% of the queries.
This is probably due to a similar effect for access path selection, as INLJs are represented in the training set as beneficial.
We additionally report the pick rate and selected runtime in \Cref{fig:physical_operator_breakdown}, which shows similar observations.
\subsection{Additional Indexes on Filter Columns}
\begin{figure*}
    \centering
    \includegraphics[width=1\linewidth]{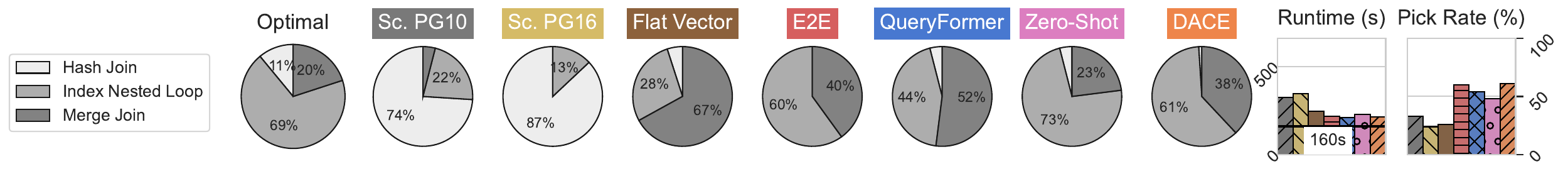}
    \caption{Breakdown with additional indexes on the filter columns on the IMDB test queries.}
\label{fig:physical_operator_breakdown_idx}
\end{figure*}

In the previous experiments, only indexes on the primary keys are used. 
In addition, we now add indexes on the filter columns, which open up additional ways of using INLJs because they also enable faster look-ups for non-primary key columns.
This shifts the distribution of which physical operators are optimal, as we discuss next.
For this, we repeat the previous experiment with additional indexes on the filter columns and show the breakdown in \Cref{fig:physical_operator_breakdown_idx}\footnote{As \qppnet cannot predict the cost for additional indexes due to its fixed featurization using one-hot encoding, we excluded it from this experiment.}.
As we can see, the use of INLJ when selecting operators optimally increases to 69\%. 
Interestingly, the performance of \postgresx and \postgresxvi drops significantly as it still prefers HJ (74\% and 87\%), and it is now outperformed by most \lcms in terms of runtime and pick rate.
Thus, it seems that the cost model of PostgreSQL is not well calibrated and still prefers hash operations.
On the other hand, the benefit of \lcms might stem from their over-preference for INLJ, which, in this experiment, luckily, is often the better choice.

\subsection{Summary \& Takeaways}
Overall, no cost model could achieve a near-optimal runtime when selecting physical query operators. 
Still, the classical models performed best and \dace as a database-agnostic model is close according to the selected runtime.
However, as mentioned, \dace learns from PostgreSQL costs, which is highly beneficial for \dace, as shown in our ablation study. 
Furthermore, another observation is that most \lcms showed a strong over-preference for INLJ, which supports the findings of the access path selection study that bias in training data is a potential problem one needs to tackle for \lcms. 
However, this is also non-trivial as INLJs can cause high runtimes and, thus, often timeout during training data collection (as they need hours or even days). 
Thus, an interesting avenue of research is to include negative signals in training the models without the need actually to run these costly negative examples.
\section{Recommendations for Cost Models} \label{sec:lessons}
In this paper, we comprehensively evaluated \lcms against traditional cost models for query optimization. 
In the following, we summarize our results and recommendations.
As a main outcome, in \textit{none} of the tasks we analyzed, \lcms can significantly beat traditional approaches for cost modeling -- even though \lcms provide higher accuracy over query workloads.
The total execution time of plans selected by the \lcms in query optimization was, in fact, often even higher than with traditional models.
However, we still believe that \lcms have a high potential, but the focus \textit{only} on their accuracy has resulted in the position we are in today.
In the following, we distill recommendations based on our main findings to provide future directions to unlock the full potential of \lcms.

\noindent \textbf{R1: Consider Model Architectures and Features.}
As shown in the classification of models in \Cref{subsec:taxonomy}, \lcms largely vary in input features, query representation and model architecture.
We summarize the most critical learnings:
(1) Learning from the query plan is absolutely necessary, but using the SQL string as input alone is unsuitable
(2) Simple model architectures like \flatvector often perform relatively well, making it questionable whether very complex architectures are necessary. 
(3) DB-agnostic \lcms often outperformed DB-specific models, as they were trained on a larger variety of query workloads and data distributions.
(4) While precise cardinality estimates help solve the downstream tasks, it remains unclear to what extent statistics and sample bitmaps help. 

\noindent \textbf{R2: Use Appropriate Metrics.}
Another key finding of this paper is that traditional evaluation strategies are insufficient to assess how good \lcms are for query optimization. 
Therefore, we recommend using the metrics presented in this paper that show how well an \lcm selects from multiple plan candidates, how well it can rank these plans, and what speed-up it provides for a given query optimization task.
Ideally, these metrics influence the design and learning approach of future \lcms.
For example, one direction could be to apply ranking-based approaches, which have been used recently for end-to-end learned optimizers \cite{behr2023, chen2023, zhu2023} to cost models.

\noindent \textbf{R3: Diversify Your Training Data.}
The third key finding of this study is that traditional training strategies for \lcms induce fundamental biases stemming from how training data is collected.
Specifically, \lcms are typically trained on \textit{pre-optimized queries}, as these have already been executed by a database system  (cf. \Cref{fig:learning_procedure}\circles{B}) to generate the corresponding labels.
Another reason for biases in training data is timeouts, which are necessary to execute training queries within a limited, reasonable time.
Query plans with expensive operators like nested-loop joins are thus likely to timeout before completion. 
Consequently, only the cases where these operators are beneficial are included in the training data, leading to a bias in the models, as for access path selection (\Cref{sec:access_path_selection}). 
A similar bias was observed for learned cardinality estimation in \cite{reiner2023}.
Overall, the traditional training strategies are thus particularly ill-suited for query optimization, as they lead to a substantial divergence between the training data distribution and non-optimized queries that occur during query optimization.
To resolve this bias, training data should be diversified so that both good and bad query plans can be learned from.

\begin{figure}
    \centering
    \includegraphics[width=0.9\linewidth]{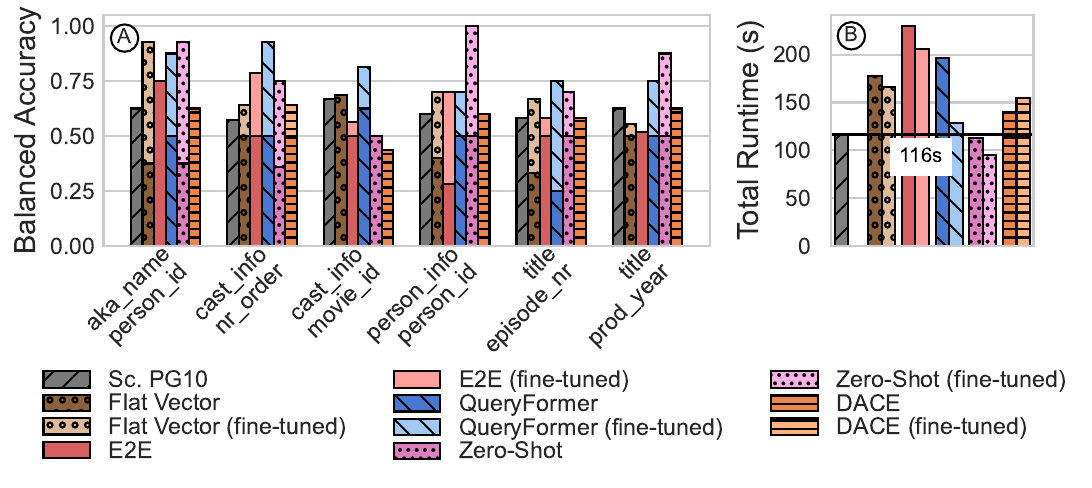}
    \caption{Fine-tuning for access path selection on IMDB columns. 
    \circles{A} For most columns, balanced accuracy increases after fine-tuning.
    \circles{B} Also the total runtime improved, so that \zeroshot outperforms \postgres.}
    \label{fig:retraining}
\end{figure}

We demonstrate the effect of how diversification can help with the task of access path selection in a small experiment.
In particular, we fine-tune \lcms with additional training data consisting of 500 randomized scan queries with different selectivities.
For each query, we enforce two executions: One with an \texttt{IndexScan} and one with a \texttt{SeqScan}, and we do not enforce timeouts.
We report the balanced accuracy $B$ over the columns from IMDB of \Cref{tab:scan_costs_over_datasets} \textit{before} and \textit{after} fine-tuning in \Cref{fig:retraining}~ \circles{A}.
As we can see, using diversified training data is highly beneficial in most cases as accuracy improves.
Even more importantly (see \Cref{fig:retraining}~\circles{B}), this also leads to improvements of up to 45\% in the total runtime across all \lcms.
Interestingly, \zeroshot is the only cost model that outperforms the runtime of \postgresx (95s vs. 116s), which shows that \lcms can outperform traditional approaches for downstream tasks and training data plays a crucial role.
The results for \dace typically do not improve because it seems to learn mainly from PostgreSQL costs as a signal, as discussed in \Cref{subsec:access_path_selectivity}.

However, while this diversification of training data has already been leveraged in the domain of learned query optimization (e.g., by cardinality injection \cite{zhu2023, doshi23} or exploration \cite{yang_balsa_2022}), its application for \lcms is not trivial.
One challenge is that each query plan would result in many possible candidates (e.g., join order permutations).
Sophisticated strategies are required to select meaningful training queries to maximize information gain and still keep the training data execution costs reasonable.
Related promising ideas are data-efficient training strategies \cite{agnihotri2024}, pseudo-label generation \cite{liu2022}, geometric learning \cite{reiner2023}, or simulation \cite{yang_balsa_2022}.
Another challenge is extremely long-running query plans, which cannot be executed in a practical sense as one plan would take days.
For this, other strategies are needed to reflect also timeout queries in the training procedure.
This is again non-trivial, as simply using a high constant value representing timeout queries leads to difficulties, as the \lcm cannot really learn meaningful costs from this.

\noindent \textbf{R4: Do Not Throw Expert Knowledge Away.}
We observed that using estimates from PostgreSQL is beneficial for \lcms, as demonstrated by the comparably good results of \dace and \qppnet.
Such a hybrid approach combines both the expert knowledge incorporated in traditional approaches and the strength of ML, which can learn arbitrarily complex functions.
While \dace and \qppnet inherently use these estimates as training features, recent work explicitly combines traditional cost functions with learned query-specific coefficients as a hybrid approach \cite{yang2023}.
This hybrid approach is particularly promising when looking at situations where \lcms still face major challenges, and traditional models are more reliable, like string-matching operations or user-defined functions.
\section{Related Work}\label{sec:related_work}
In the following, we structure and discuss related work.

\textbf{Learned Cost Models (\lcms).}
As discussed in \Cref{sec:introduction}, existing works on \lcms do not evaluate against query optimization tasks.
Only a few evaluations exist, that in contrast to our evaluation lack either the consideration of diverse query optimization tasks or analyze only a limited number of learned approaches.
For example, an early analysis \cite{wu2013} also compares traditional to ML-based cost models. 
However, only simple ML approaches were examined.
Moreover, \cite{leis_how_2015} analyzes the quality of cardinality estimates and cost models of traditional query optimizers, but they do not evaluate learned approaches.
Another line of work evaluates individual aspects, such as the impact of different query plan representation methods and featurizations on cost estimation \cite{zhao2024, chang2024}. 

\textbf{Learned Cardinalities (LCE).}
The idea of learning cardinalities has been widely studied \cite{kipf2019, hilprecht2020deepdb, naru2019, yang2020} and often evaluated.
\cite{wang2021} showed that LCE approaches are often more accurate than traditional methods but come with high training and inference efforts.
\cite{sun2021} revisits learned cardinalities and proposes a unified design space.
However, neither study evaluates the effects of query optimization.
Different is \cite{kim2022}, which showed that the overall runtime using LCE approaches can significantly reduce against PostgreSQL in many cases.
Another recent work \cite{reiner2023} showed that a similar bias occurs for LCE as we show in our work for \lcms, as they typically rely on training samples that are provided by the PostgreSQL optimizer and thus near-optimal.
They address this bias by proposing a novel geometric LCE approach that requires fewer training samples.
However, these approaches cannot be easily transferred to \lcms.

\textbf{Learned Query Optimization (LQO).}
The idea of LQO is to directly predict an optimal query execution plan or execution hints given a SQL query without the use of a cost model \cite{marcus_bao_2021, yang_balsa_2022, chen2023, zhu2023}.
While these approaches improve the overall query performance, the generality of these results is limited, as shown by \cite{lehmann2024}.
Their work found that PostgreSQL still outperforms recent LQO approaches in many cases, especially for the end-to-end execution time, including inference and plan selection.
Similar to our work, they point out that learned database components often do not behave as expected and highlight biased evaluation strategies.
However, they focus on reinforcement learning approaches and the effect of different sampling strategies for obtaining splits of the training and test data.
\section{Conclusion} \label{sec:conclusion}
In this paper, we analyzed how good recent \lcms really are for query optimization tasks.
Particularly, we experimentally evaluated seven recent \lcms for join ordering, access path selection, and physical plan selection.
Although \lcms are basically capable of learning complex cost functions, they are often inferior to traditional approaches in selecting possible execution plans.
To improve future \lcms for query optimization, we recommend using appropriate metrics, diversification of training data, and hybrid models that incorporate estimates from traditional cost models.
\section*{Acknowledgements}
This work has been supported by the LOEWE program (Reference III 5 - 519/05.00.003-(0005)), hessian.AI at TU Darmstadt, the IPF program at DHBW Mannheim, as well as DFKI Darmstadt.
\bibliographystyle{ACM-Reference-Format}
\bibliography{bibliography.bib}
\end{document}